\documentclass{ptephy_v1}
\preprintnumber{XXXX-XXXX} 

\usepackage{url} 

\usepackage{ulem}

\begin{document}
\title{Overview of KAGRA: Calibration, detector characterization, physical environmental monitors, and the geophysics interferometer
}


\author[1,2]{T.~Akutsu}
\author[3,4,1]{M.~Ando}
\author[5]{K.~Arai}
\author[5]{Y.~Arai}
\author[6]{S.~Araki}
\author[7]{A.~Araya}
\author[3]{N.~Aritomi}
\author[8]{H.~Asada}
\author[9,10]{Y.~Aso}
\author[11]{S.~Bae}
% ...

\affil[1]{Gravitational Wave Science Project, National Astronomical Observatory of Japan (NAOJ), 2-21-1 Osawa, Mitaka City, Tokyo 181-8588, Japan}
\affil[2]{Advanced Technology Center, National Astronomical Observatory of Japan (NAOJ), 2-21-1 Osawa, Mitaka City, Tokyo 181-8588, Japan}
\affil[3]{Department of Physics, The University of Tokyo, 7-3-1 Hongo, Bunkyo-ku, Tokyo 113-0033, Japan}
\affil[4]{Research Center for the Early Universe (RESCEU), The University of Tokyo, 7-3-1 Hongo, Bunkyo-ku, Tokyo 113-0033, Japan}
\affil[5]{Institute for Cosmic Ray Research (ICRR), KAGRA Observatory, The University of Tokyo, 5-1-5 Kashiwa-no-Ha, Kashiwa City, Chiba 277-8582, Japan}
\affil[6]{Accelerator Laboratory, High Energy Accelerator Research Organization (KEK), 1-1 Oho, Tsukuba City, Ibaraki 305-0801, Japan}
\affil[7]{Earthquake Research Institute, The University of Tokyo, 1-1-1 Yayoi, Bunkyo-ku, Tokyo 113-0032, Japan}
\affil[8]{Department of Mathematics and Physics, Hirosaki University, 3 Bunkyo-cho, Hirosaki City, Aomori 036-8561, Japan}
\affil[9]{Kamioka Branch, National Astronomical Observatory of Japan (NAOJ), 238 Higashi-Mozumi, Kamioka-cho, Hida City, Gifu 506-1205, Japan}
\affil[10]{The Graduate University for Advanced Studies (SOKENDAI), 2-21-1 Osawa, Mitaka City, Tokyo 181-8588, Japan}
\affil[11]{Korea Institute of Science and Technology Information (KISTI), 245 Daehak-ro, Yuseong-gu, Daejeon 34141, Korea}
% ...

\date{\today}
\begin{abstract}%
KAGRA is a newly built gravitational wave observatory, a laser interferometer with a 3 km arm length, located in Kamioka, Gifu, Japan. 
In this series of articles, we present an overview of the baseline KAGRA, 
for which we finished installing the designed configuration in 2019. 
This article describes the method of calibration (CAL) used for reconstructing gravitational wave signals 
from the detector outputs, as well as the characterization of the detector (DET).
We also review the physical environmental monitors (PEM) system and the geophysics interferometer (GIF). 
Both are used for characterizing and evaluating the data quality of the gravitational wave channel. 
They play important roles in utilizing the detector output for gravitational wave searches. 
These characterization investigations will be even more important in the near future,
once gravitational wave detection has been achieved, and in using KAGRA in the gravitational wave astronomy era.

\end{abstract}

\subjectindex{xxxx, xxx}

\maketitle
\section{Introduction}\label{Sec:main:intro}

Gravitational wave (GW) astronomy is becoming
one of the most exciting research fields in physics and related disciplines. 
Since the first direct detection of GWs from a binary black hole merger \cite{GW150914PRL}, 
many GW signals have been detected by the LIGO \cite{aligo} and Virgo \cite{adV} interferometers.
Moreover, the first detection of a GW signal from a binary neutron star merger in 2017 \cite{GW170817PRL}
has opened the era of multi-messenger astronomy \cite{mmaGW170817}.

KAGRA \cite{qnd} is a GW interferometer located in Japan.
It is termed a 2.5$^{th}$-generation GW interferometer 
because it is constructed underground \cite{iKAGRA:PTEP} 
and operated at cryogenic temperatures (20K) \cite{bKAGRAphase1}
Underground construction and cryogenic operation
are essential techniques for the next-generation detectors \cite{ET, CE}. 
By April of 2019, the installation work was mostly
completed, and the interferometer was commissioned \cite{KAGRA:PTEP01}.
At the end of August 2019, the first interferometer control with Fabry-Perot Michelson (FPMI) configuration was established, 
and by the end of January 2020, a power-recycled Fabry-Perot Michelson interferometer (PRFPMI) configuration had been established.
Finally, the GEO600 \cite{geo} and KAGRA interferometers 
conducted a joint two-week observation run, called ``O3GK'', in April 2020.

Preparation of the calibration instruments and 
understanding the characterization of the interferometer 
play important roles in the accurate reconstruction of strain.
Reducing the systematic errors in GW signal reconstruction with lower bias leads to 
accurate GW parameter estimation. 
Precise mass evaluation from compact binary coalescences (CBCs) provide information about the  
origin of the binary and the evolution of the universe.
Precise spatial identification of a GW source in the sky (sky localization)
provides a wealth of knowledge for multi-messenger astronomy and 
allow identification of the host galaxy.

Detector characterization plays an essential role in distinguishing a GW signal from detector noise.
Unexpected behavior of the interferometer causes transient noise, which can make false detection of gravitational waves.
Detector characterization plays a role in identifying such interferometer status and noise behavior.
The identification results shows whether to use the data for the gravitational wave
search or discard it as a false event.
It reduces the false alarm rate and improves the SNR of GW signals.
Not only main interferometer signal but also auxiliary channels are used in order to evaluate the noise behavior.
Auxiliary sensors in optics and vibration isolated system are useful for investigating noise derived from control of the interferometer.
During the O1 and O2 observations\cite{O1O2catalog}, only GWs from CBCs were successfully detected.
Understanding the origin of detector noise is critical for data analysis of GWs searches and parameter estimation;
detection of GW signals from \textit{new} sources may significantly expand our knowledge.
Also, to identify noise derived from external disturbances, physical environmental monitorings (PEMs) are essential.
To identify noise due to disturbances external to the detector, physical environmental monitorings (PEMs) are essential.
Selection of used data and rejection of candidate events by using interferometer behavior and the information of identified noise worked well in past observations (O1 and O2) by LIGO and Virgo\cite{DET:ref_ligoveto,DET:ref_ligoglitch}.
As a result, they achieved 67 detection of GWs from CBCs\cite{O1O2catalog}.

A geophysics interferometer (GIF) was constructed in the 
KAGRA X-arm tunnel and has been operating continuously since 2016.
The 1,500m GIF provides precise measurements of ground motions in the 
underground environment, which also can be used in the KAGRA arm-length-compensation system.

Section \ref{Sec:main:CAL} summarizes the history of the KAGRA calibration activity.
Section \ref{Sec:main:DET} discusses the data acquisition/quality and
transient noise identification.
Section \ref{Sec:main:PEM} describes the introduction history of the KAGRA PEMs and highlights some of their controbutions.
Section \ref{Sec:main:GIF} provides a description of the geophysics interferometer and its installation, and detector performance
improvements it has enabled.
Section \ref{Sec:main:Conclusion} summarizes this paper.

\section{Calibration}\label{Sec:main:CAL}
\subsection{Introduction}\label{Sec:CAL:intro}
\begin{figure}
\begin{center}
\includegraphics[width=12cm]{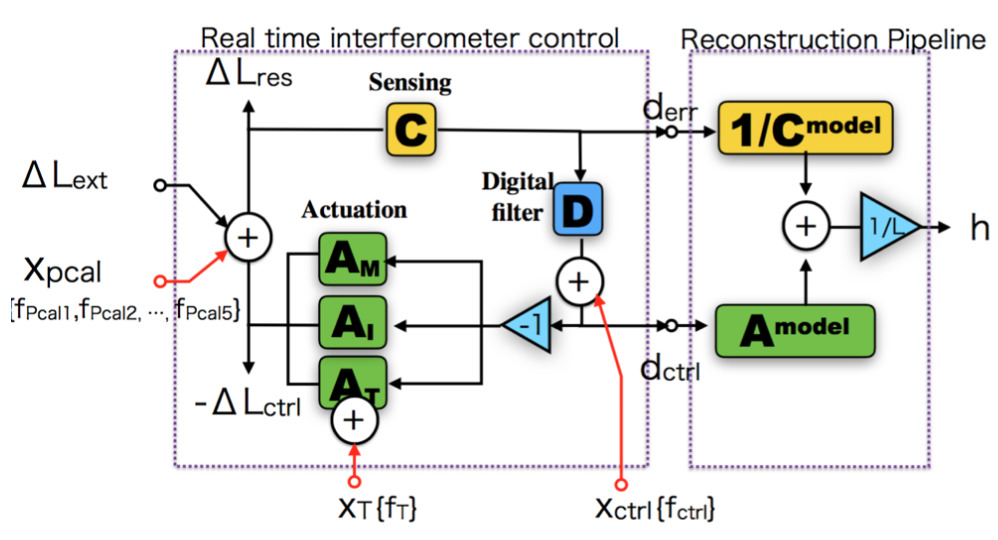}
\caption{Diagram of the DARM control loop and the reconstruction pipeline. 
The DARM model consists of the actuation and sensing parts. 
The actuator part corresponds to the transfer function from a digital-to-analog converter (DAC) to the displacement of the mirror. 
A$_M$, A$_I$, A$_T$ are the actuator efficiency of the 
marionette stage, intermediate mass stage and test mass stage, 
respectively.
The sensing part is a combination of interferometer and photo-detector responses. 
The quantities $d_{err}$ and $d_{ctrl}$ are error and control signals, which are outputs from the interferometer. 
f$_{Pcali}$ is the injection frequency by photon calibrator.  
Using this DARM model, we can construct actuator part and sensing part for reconstructing the signal.
Red arrows show injections from outside the feedback loop. }
\label{fig:DARM_diag}
\end{center}
\end{figure}

The main purpose of the calibration studies is to provide the strain and its error~\cite{LIGO:CAL:opt,CAL:Evan,CAL:Craig}. 
Development of the calibration instruments and reconstruction pipelines is essential for precise calibration of the detector. 
In this process, we need to consider the relationship between the strain and a model of the GW detector. 
The time series (${\it h}(t)$) of GW strain is given by
\begin{equation}
{\it h}(t)=\frac{\Delta L_{ext}(t)}{L_0}=\frac{L_{x}(t)-L_{y}(t)}{L_{0}},
\end{equation}
where $L_0$ is the effective length of a KAGRA arm (3,000 m) and 
$\Delta L_{ext}(t)=L_x(t)-L_y(t)$ is the difference between the x- and y-arm lengths
, as described in Fig. \ref{Fig:PEM:PEMMAP}, 
caused by external sources; 
the strain {\it h}($t$) is not directly available from the interferometer output.
The KAGRA interferometer is controlled by digital feedback loops. 
Four length-control loops
--the Michelson differential length (MICH), power-recycling cavity length (PRCL), common-mode arm cavity length (CARM) and differential arm length (DARM)--
were used for control during O3GK~\cite{CAL:Michimura}.
A model diagram of the KAGRA DARM feedback loop is shown in Fig.~\ref{fig:DARM_diag}. 
In this diagram, the model consists of a real-time interferometer-control part and a reconstruction-pipeline part. 
Real-time interferometer control is based on sensing and actuation functions, together with digital filter. 
The sensing function corresponds to the optical response of the interferometer and its readout, 
and the actuation function corresponds to the efficiency of the coil-magnet actuator on the end test mass. 
The digital-filter is a component of the real-time control system. 
This model enables us to use an analytic formula to calculate the transfer functions. 
However, the sensing and actuation functions include time-dependent parts~\cite{CAL:Darkhan}. 
A set of measurements of the sensing and actuation functions is thus necessary to complete the DARM model.
The external displacement, $\Delta L_{ext}(t)$, is calculated from the digital signals $d_{err}(t)$ and $d_{ctrl}(t)$. 
Using Fig.~\ref{fig:DARM_diag}, we obtain
\begin{eqnarray}
\Delta L_{res}(t)&=&\Delta L_{ext}(t) -\Delta L_{ctrl}(t), \\
d_{err}(t)&=&C * \Delta L_{res}(t), \\
\Delta L_{ctrl}(t)&=&A * d_{ctrl}(t).
\end{eqnarray}
By combining with above equations, the external excitation is obtained as
\begin{equation}
\Delta L_{ext}(t) = \left[ \frac{1}{C} * d_{err}(t) \right] + \left[ A* d_{ctrl}(t)\right],
\end{equation}
where the convolution operation is defined by $F*G(t)=\int F(t')G(t-t')dt'$, $F(t)$ is a time-domain filter, and $G(t)$ is a digital signal. 
Precise calibration is required to measure the actuation and sensing functions accurately.

\subsection{DARM model}\label{Sec:CAL:model}
The calibration instruments provide the parameter information needed to determine the DARM model precisely through the response function $\tilde{R}$~\cite{CAL:Craig}, 
which is defined as follows:
\begin{equation}
\Delta \tilde{L}=\tilde{R}\tilde{d}_{err}=\left( \frac{1+\tilde{G}}{\tilde{C}} \right) \tilde{d}_{err},
\end{equation}
where the open loop gain is $\tilde{G}=\tilde{C}*\tilde{D}*\tilde{A}$.
The response function is thus given by the following equation:
\begin{equation}
\tilde{R}(f,t)=\frac{1}{\tilde{C}(f,t)}+\tilde{D}(f)\tilde{A}(f,t),
\end{equation}
where $\tilde{A}(f,t)$ and $\tilde{C}(f,t)$ are models of the actuation and sensing functions. These function is defined as:
\begin{eqnarray}
A(f,t)&=&\Sigma_i^{\{M,I,T\}}H^{(i)}_a A^{(i)}(f)exp(-2\pi f i \tau^{(i)}_a), \\
C(f,t)&=&\frac{H_c}{1+f_c/f}C(f)\exp(-2 \pi f i \tau_c)
\end{eqnarray},
which {$M$,$I$,$T$} are the marionette stage, 
intermediate mass stage and test mass stage, respectively.
To complete the calibration model, the parameter set $\vec{\theta}=\{ H_c,f_c,\tau_c,H^{(i)}_a,\tau^{(i)}_a \}$ is measured using a swept-sine injection test, 
where $H_c$, $f_c$, and $\tau_c$ correspond to the optical efficiency, cavity pole frequency,
and time delay of the sensing function, and $H^{(i)}_a$ and $\tau^{(i)}_a$ are the actuation efficiency and time delay from the i-th suspension mass, respectively. 
The Markov-Chain Monte Carlo (MCMC) method is used to determine $\vec{\theta}$ based on the swept sine measurements from the coil magnet actuator and photon calibrator. 
The MCMC algorithm provides posterior probability distributions of the model parameters, 
with a likelihood $L(M,\vec{d}|\vec{\theta})$ and an assumed prior distribution. 
The likelihood is defined using least-squares minimization between the model $M$ and the measured data $\vec{d}$.
The parameters so-determined are also used in each reconstruction pipeline.

\subsection{Calibration instruments}\label{Sec:CAL:instrument}
We have developed both a photon calibrator (PCAL)~\cite{CAL:Karki} and a gravity field calibrator (GCAL)~\cite{CAL:Yuki1} for precise calibration of the GW detector. 
They allow us to determine the actuation and sensing functions and complete our DARM model. 
To calibrate the sensing and actuation functions, a displacement has to be applied to produce a differential change in the arm length. 
Classically, the free-swinging Michelson method has been used to calibrate displacements. 
It uses the wavelength of a laser as the length standard. 
However, calibration using photon pressure is a modern method used today \cite{CAL:LIGOcal,CAL:LIGOPCal}
and one using gravity fields is being studied and developed for more precise calibration in future observations \cite{CAL:Yuki1,CAL:Extevez}.
Figure~\ref{fig:gpcal} shows an overview of the KAGRA calibration instruments. 
The PCAL was used as the main calibrator of the KAGRA observatory during O3GK.  
The KAGRA PCAL was placed 36 m away from the end test mass, 
and a stabilized laser beam was injected with selected frequencies 
onto the mirror surface to produce a displacement. 
During O3GK, we selected three frequencies.
We also plan to install gravity-field calibrators at the front of the end test masses~\cite{CAL:Matone,CAL:Matone2,CAL:Matone3}. 
A gravity-field calibrator generates a gravity-field gradient around the end test mass. 
By calculating the force exerted by a quadrupole mass distribution, we can determine the motion of the test mass very accurately.

\subsubsection{Photon calibrator}\label{Sec:CAL:pcal}
The PCAL was originally developed at the GEO600 and Glasgow 10 m interferometers 
and is regarded as a 1$^{st}$-generation photon calibrators~\cite{CAL:Clubley,CAL:Mossavi}. 
By using photon pressure, the investigators succeeded in actuating the mirror surface. 
However, they reported elastic deformations at the injection points,
which were the  centers of mass of mirrors~\cite{CAL:Hild}. 
To avoid elastic deformations, LIGO developed a 2$^{nd}$-generation PCAL system that uses two-point injections~\cite{CAL:Karki}, 
which move the node of the drum-head mode to mitigate elastic deformations~\cite{CAL:Evan}. 
An optical-follower servo was also developed to reduce laser noise and higher harmonics.
In this paper, we discuss a 3$^{rd}$ generation PCAL for KAGRA, which was developed by a collaboration between KAGRA and LIGO.  
To understand the high-frequency response, a 20 W continuous-wave laser is used with an optical-follower servo, 
and the operating power was increased to be 10 times larger than that of a 2$^{nd}$ generation system. 
An independent beam-control system was also employed to characterize the response of the test-mass pendulums. 
Monitoring of the beam position is also necessary to characterize rotation and elastic deformations. 
Telephoto camera which is described in Fig. \ref{fig:gpcal}, 
can monitor the beam position.
The response from the detector is converted into power using a laser-power standard calibrated by the National Institute of Standards and Technology (NIST)~\cite{CAL:NIST}. 
We calibrate the PCAL response every month. The relative uncertainty in laser power obtained from the laser-power standard measurement is 0.32 \%. 
However, the absolute laser power has a 3 \% uncertainty, because it is determined by the absolute power measurement based on the NIST power standard, 
and the power standard of each country has a variance of 3 \%. 

\subsubsection{Gravity-field calibrator}\label{Sec:CAL:gcal}
\begin{figure}
\begin{center}
\includegraphics[width=14cm]{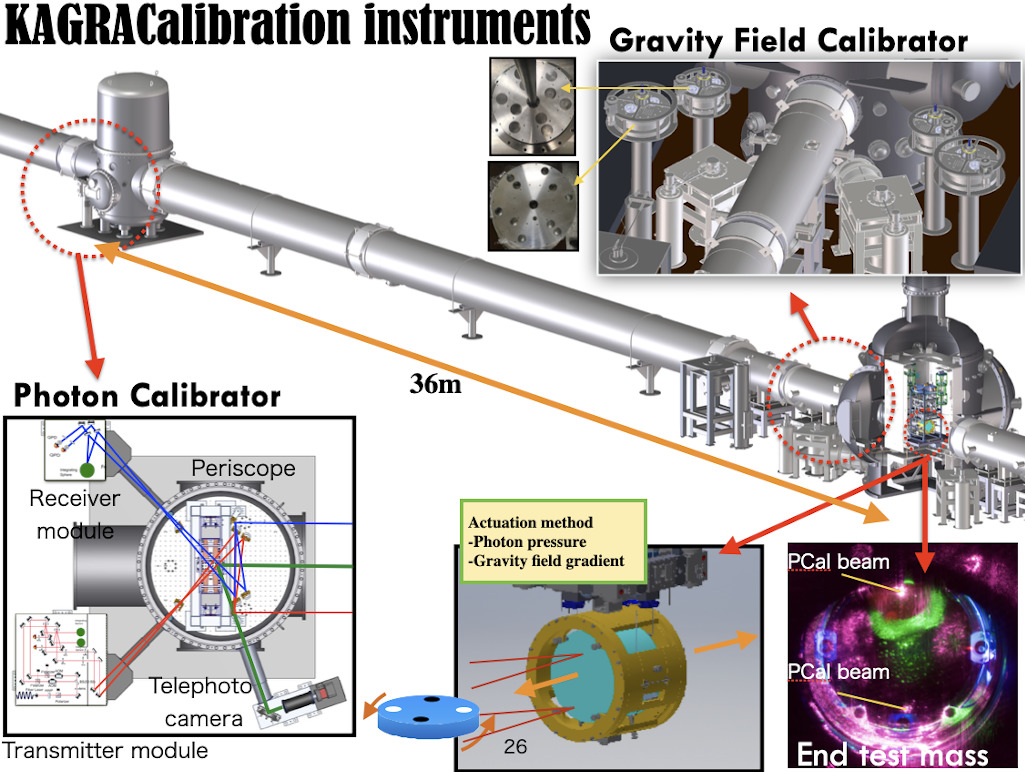}
\caption{ Schematic view of the KAGRA calibration instruments. 
The photon calibrator is placed 36 m from the end test mass. 
Beams from the transmitter module are injected onto the mirror surface to it. 
The expected displacement is monitored by using a read-back signal at the receiver module. 
The gravity-field calibrators are installed around the end test mass. 
The gradient of the gravity field changes the position from the test mass. 
The expected displacement is calculated from the masses of the rotors and the geometry.
To monitor the injected beam position, we also installed the Telephoto camera, which is the combination of telescope and photo-camera. Telephoto camera can monitor not only the PCal beam position, but also the main beam position and surface of sapphire mirror.
}
\label{fig:gpcal}
\end{center}
\end{figure}

The GCAL is a new type of calibrator for absolute calibration. 
When we calibrate the interferometer response using the PCAL, 
the absolute error in laser power is propagated directly into the uncertainty in the gravitational-wave strain. 
To avoid this problem, we have newly developed a gravity field calibrator system for KAGRA. 
The original design for this calibrator is based on the CLAB experiment at KEK and the University of Tokyo~\cite{CAL:Hirakawa,CAL:Oide,CAL:Suzuki,CAL:Ogawa,CAL:Kuroda}. 
We took over the system design of the previous experiment and improved it with current technology; 
the original design was tested 40 years ago. 
We replaced the motor, encoder, and vacuum seal with state-of-the-art designs. 
Virgo developed the same concept independently, which they called a 
``Newtonian calibrat''or~\cite{CAL:Extevez}. 
Virgo performed a demonstration to measure the displacement by them.
We will employ the new system for collaborative worldwide observations. 
The KAGRA gravity-field calibrator system consists of four subsystems. 
As shown in Fig.~\ref{fig:gpcal}, the gravity-field calibrators are placed at the left and right sides of the chambers for symmetry. 
The left and right calibrators cancel the systematic errors due to rotation. 
Large and small calibrators are used for consistency checks of the displacement. 
To verify the model uncertainty, we cross-check the expected response of the mirror using both large and small rotors. 
Four rotors are synchronized using a rotary encoder and its readout system. 
By monitoring the rotation, we can determine the expected displacement~\ref{CAL:Carbone1}. 
At the same time, we need to monitor the absolute distance between the center of the GCAL and the position of the end test mass. 
By using the hexapole distribution of the rotor, we can cancel systematic errors in the absolute distance measurement.

\subsection{Reconstruction pipelines}\label{Sec:CAL:pipeline}
Three types of pipelines will develope to calculate the GW strains, called the C00, C10, and C20 pipelines. 
Each pipeline has its own characteristics. 
The purpose of each pipeline is explained below.
\subsubsection{C00: The online pipeline}\label{Sec:CAL:C0pipeline}
The main purpose of the C00 pipeline is to monitor {\it h}($t$) during the operation of the interferometer. 
It is a online calibration pipeline that employs infinite-impulse response (IIR) filtering techniques. 
Using the output of the online system, we multiply the actuation and sensing function models by the IIR filters. 
We update these parameters every week. 
We neglect time dependence in this pipeline. 
By using the IIR filter, we approximate the high-frequency response as a time delay effect.

\subsubsection{C10: Low-latency pipeline}\label{Sec:CAL:C1pipeline}
\begin{figure}
\begin{center}
\includegraphics[width=14cm]{ 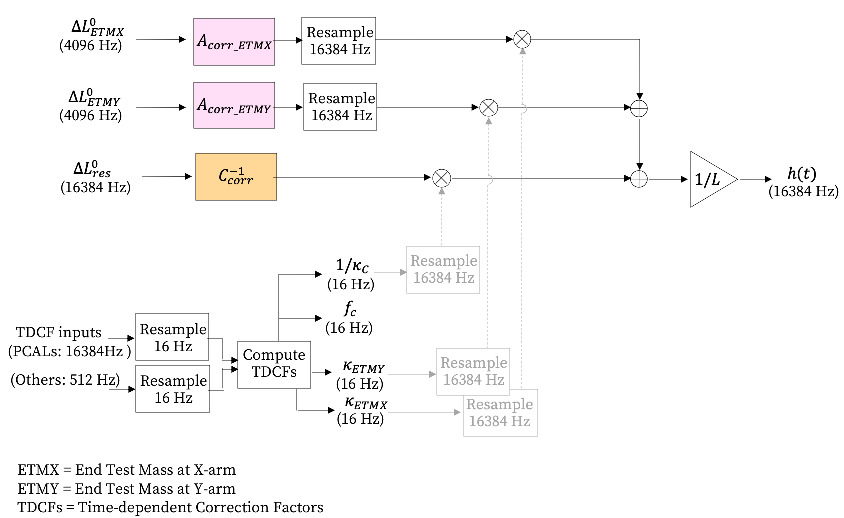}
\caption{Conceptual diagram of the low-latency calibration pipeline. 
The partially-calibrated outputs of the front-end calibration pipeline,
$\Delta L_{ETMX}^0$, $\Delta L_{ETMY}^0$ and $\Delta L_{res}^0$, 
are used as inputs. 
They are filtered by the FIR correction filters in the actuation and inverse sensing paths, added together
with time-dependent correction factors(TDCFs, under discussion), and then divided by $L$
to give the strain signal $h(t)$.}
\label{fig:lowlatency}
\end{center}
\end{figure}

The main purpose of the C10 pipeline is for low latency analysis. 
This pipeline receives DARM loop signals that are partially calibrated with the IIR in the C00 pipeline as shown in Fig.~\ref{fig:lowlatency}. The time dependent factors are also monitored with calibration lines.
The C10 reconstruction filters are calculated with appropriate finite-impulse response (FIR) filters using a GStreamer-based pipeline known as ``gstlal''~\cite{CAL:gstlal}. 
This pipeline will generate {\it h}($t$) with a latency less than 10 seconds. 
By using an FIR filter and the demodulation signals from the calibration lines, 
the uncertainty in {\it h}($t$) can be reduced below that of obtained from C00~\cite{CAL:Darkhan}. 
This {\it h}($t$) will be used for event-search analysis for follow up telescopes and detectors, 
for which C10 pipeline generates information about the calibration status, providing calibration flags at the same time. We will also update these parameters every week.

\subsubsection{C20: High-latency pipeline}\label{Sec:CAL:C2pipeline}
The C20 calibration pipeline is also based on the gstlal. 
It produces {\it h}($t$) with offline raw data on a high-latency server. 
The high-latency pipeline will be produced with data several months 
after the acquisition of raw data. 
For this pipeline, we will adopt FIR filtering with direct error and control signals. 
The time dependence from the PCAL will also be applied in this process.  

\subsection{Error estimation}\label{Sec:CAL:error}
Error estimation for the response functions is one of the most challenging topics, because the reconstruction process is non-linear. 
Even if we attempt to fit the data, it is sometimes mismatc
hed under the linear regression~\cite{CAL:Craig}.
Gaussian-process regression (GPR) is a method of Bayesian model estimation for a non-linear system. 
In a Gaussian process, the set of data is modeled as a simple Gaussian distribution $N[m(f),\sigma(f)]$.  
The GPR results yield a distribution function around the mean of the data, which provides an uncertainty estimate at the same time. 
To apply the GPR method, we will determine the residual response function as follows:
\begin{equation}
\frac{\delta \tilde{R}(f)}{\tilde{R}_{model}(f)}=\frac{\tilde{R}_{meas}(f)-\tilde{R}_{model}(f)}{\tilde{R}_{model}(f)},
\end{equation}
where $\tilde{R}_{model}(f)$ and $\tilde{R}_{meas}(f)$ based on the parameters determined with the MCMC method and the measured response function. 
The frequency-domain response-function $\delta R$ is proportional to the GW strain error $\delta h$, as shown in eq. (9). 
We can also define the corresponding uncertainties $\sigma_{R}$ and $\sigma_{h}$ as in eq. (10)
\begin{eqnarray}
\frac{\delta \tilde{R}}{\tilde{R}^{model}} &=& \frac{\delta h}{h}, \\
\frac{\sigma_R}{\tilde{R}^{model}} &=& \frac{\sigma_h}{h}.
\end{eqnarray}
The response function must therefore be characterized in order to perform the calibration.
Finally, we will obtain the mean and uncertainty of the response function 
and determine the time-dependent errors. 
By using the DARM-model parameters parameters and time-dependent correction factors from the reconstruction pipeline, 
we will estimate the uncertainty with a Monte Carlo simulation. 
The PCAL uncertainty, based on the power calibration of the read-back signal, will be also included in the error estimation.

\section{Detector Characterization}\label{Sec:main:DET}
\subsection{Data acquisition}\label{DET:ssec_DAQ}
KAGRA is composed of nineteen suspended mirrors and many optical components \cite{KAGRA:PTEP01}.
All mirrors and optics are controlled by a digital control system.
The data-acquisition system is integrated into this digital control system, 
and it records more than 100,000 channels.
The recorded channels contain not only the main interferometer signals but also signals from physical environmental sensors, 
many test points in the control loop of the main interferometer, local control signals from all the suspended mirrors, and so on.
All data are recorded as discrete time-series signals with various sampling intervals.
The total data rate reached 12 MB/s during the O3GK observing run on KAGRA.
\textcolor{black}{This data set was obtained at the Kamioka site and was transferred to KAGRA's main data center at Kashiwa.}
The KAGRA data is distributed from Kashiwa to many computer centers located both in Japan and at overseas sites, 
including the  computer centers of LIGO and Virgo.
Details of the data transfer from KAGRA are discussed in \cite{KAGRA:PTEP04}

By using these signals, the recorded data are classified into two categories. 
One is used for scientific purposes such as searching for GWs and determining the parameters of GW sources.
The other is used solely for evaluating the detector and its noise status.
For real-time GW searches, it is difficult to analyze all channels due to limited computer resources.
However, analyzing auxiliary channels can tell us whether the quality of the interferometer data is sufficient for GW searches and parameter estimation.
For this reason, basic criteria are set for many auxiliary channels.
Some indicators, called ``data-quality state vectors'', are provided if these criteria are satisfied.
\textcolor{black}{This process is performed by the digital control system and the vectors are recorded in bit-string format.}
The process of data-quality evaluation is also performed offline in order to correct for mistakes and errors in the real-time process.
Because the amount of data reaches 1PB/yr for each detector, it's difficult to transfer all data between overseas.
From the view point of data storage, there is no enough storage for keeping all KAGRA, LIGO and Virgo's data.
Therefore only some important channels are shared with overseas.
In order to reduce the amount of data, only the main interferometer signals and the data-quality (DQ) state vector are shared with LIGO and Virgo.
In addition, a list of an GPS times when glitch is occurred that define ``science segments'' in which the data can be used for searching for GW signals is provided.
The DQ state vector was used to the data selection in order to estimate duty factor, the detection range of binary neutron stars and so on. 
duty factor and the detection range is basically estimated only with the detector data which is in the science mode.
On the other hand, these quantities are also computed for the data which is not in the science mode though the interferometer is locked in order to investigate behavior of interferometer in various states.
This ``segment'' information is provided to indicate the noise status in order to evaluate the status of the interferometer.

To search for GWs reliably, it is important to reject false events from among the GW candidate events.
Each GW search pipeline evaluates false alarm probability from the background noise behavior of the GW channel.
Other auxiliary channels are not usually used in the GW search pipelines; instead, 
they are analyzed using ``glitch pipelines'' and other noise-evaluation methods.
Glitch pipelines are tools that detect bursts of excess power, used to identify transient noise.
They are applied not only to GW channel but also many auxiliary channels.
Detected transient noise in the GW channel and auxiliary channels are evaluated for coincidence and used to remove false events from GW event candidates.

\begin{figure}
\begin{center}
\includegraphics[width=16cm]{ 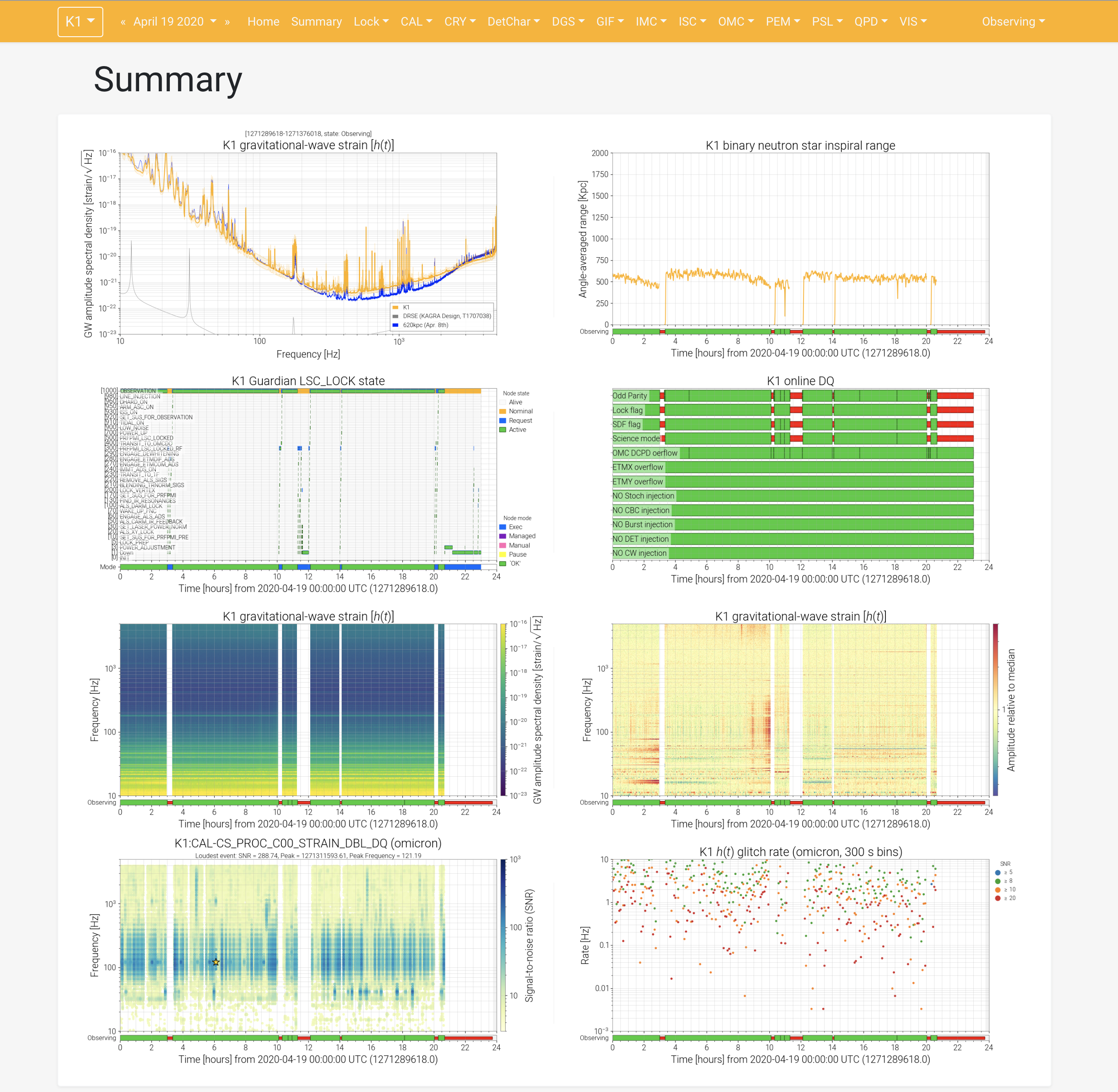}
\caption{Example of the KAGRA SummaryPages. They are used to monitor the interferometer status by on-site team members and for daily checks from remote sites.}
\label{DET:fig_summ}
\end{center}
\end{figure}

Data quality is assessed not only for reliable detection, but also for improving the sensitivity and stability of the detector.
Comparison between the current and past interferometer status often helps in finding the reason why data has been flagged as ``bad condition''.
A data-monitoring system is provided as a web interface called ``SummaryPages,'' \cite{DET:ref_gwsumm,DET:ref_gwpy} which is used to check interferometer stability and to detect changes in the 
interferometer status. 
Various plots of the main interferometer signals and many auxiliary channels are provided and archived every day.
Figure \ref{DET:fig_summ} shows an example of the KAGRA SummaryPages, which displays the latest detector sensitivity, inspiral ranges 
that indicate the detectable distance of GWs from binary neutron stars, and some data-quality flags.
The plots on the SummaryPages are updated every 15 minute and are also used for daily check from remote sites such as universities, institutes and also home of collaborators.

\subsection{Data-Quality State Vector}\label{DET:ssec_DQV}
Interferometer status is evaluated from many auxiliary channels. Because this evaluation result is used in many cases  such as the decision of interferometer control strategy, interferometer noise evaluation, various GW searches, etc., status evaluation is performed as a real-time process in automatically and its results are recorded as a simple indicator such as "OK" or "Not OK". In order to satisfy the various situations, several types of indicators were prepared during O3GK. These indicators are merged as one bit-string named as "Data Quality (DQ) state vector".
This DQ state vector helps us to use same criteria for each search pipeline and to reduce CPU costs for re-evaluation. The definition of DQ state vector is shown in Table \ref{DET:tab_DQV}.
The most important flag is the science-mode flag, because 
GW searches are performed only for data indicated to be science-mode data.
The science mode does not include any periods in which 
(1) a calibrated strain signal is not available, due to some reason such as a hang-up of the calibration process, 
(2) interferometer control setting are not nominal,  or 
(3) there are signal injections or excitation.
Periods in which saturation of analog-to-digital converters (ADCs) or digital-to-analog converters (DACs) occurs are provided as auxiliary segment information.

KAGRA's digital control system makes it easy to change the configuration, as compared with analog control systems in general.
On the other hand, managing the definition of the control settings becomes more important for providing the reliable data because changes in the configuration easily changes calibration, sensitivity, behavior of background noise, and so on.
Operating the interferometer with any different configuration changes the stability and sensitivity to GWs.
Such a change affects noise background estimation in each GW-search pipeline.
All of the interferometer control configurations were well defined as nominal settings during interferometer commissioning.
If the interferometer status changes due to time variations or some trouble, the nominal settings are set again after human validation.
Unexpected differences between the latest configuration and the nominal configuration can be detected through KAGRA's digital control system.
Any period with at least one setting different from the nominal one is flagged as a non-science mode.

The data-quality state vector also inclues injection flags.
Signal injections are performed in order to measure the interferometer response to GWs, investigate sources of detector noise that limit the sensitivity to GWs, and check the calibrated strain signals and GW-search pipelines.
For those purposes, signals are injected from coil-magnet actuators on the suspended mirrors or from the photon calibrators through the digital control system.
Measuring the response of the interferometer is performed by using sine, swept-sine, and sine-Gaussian waveforms.
Various theoretical GW waveforms are used to check calibrated strain signals and test the GW-search pipelines.
Because signals due to these injections must be excluded from candidate events, any periods with injections are flagged.
Five injection flags are provided to indicate the type of waveform being injected.
As shown in Table \ref{DET:tab_DQV}, there are five different kinds of injection flags for the different waveforms.

\begin{table}
  \begin{center}
    \caption{Definition of KAGRA data-quality state vector}
    \begin{tabular}{|c|l|} \hline 
      Bit & Meaning of flags \\\hline \hline
      0 & Odd parity \\\hline
      1 & Lock-check flag \\\hline
      2 & Control-setting check flag \\\hline
      3 & Science-mode flag \\\hline
      4 & ADC overflow \\\hline
      5 & DAC End test mass X(ETMX) overflow \\\hline
      6 & DAC End test mass Y(ETMY) overflow \\\hline
      7 & Injection flag for stochastic gravitational wave background \\\hline
      8 & Injection flag for compact binary coalescence waveform \\\hline
      9 & Injection flag for burst waveform (e.g, Supernovae) \\\hline
      10 & Injection flag for detector characterization \\\hline
      11 & Injection flag for continuous wave waveform (e.g, pulsars) \\\hline
    \end{tabular}
    \label{DET:tab_DQV}
  \end{center}
\end{table}

\subsection{Data-Quality Segment}\label{DET:ssec_DQSEG}
Segment information is generated to indicate multiple data periods that are suitable for use in gravitational wave searches.
One segment is recorded between the two GPS times when a science mode starts and ends.
In addition to the science mode, segment information is provided about overflow periods and various types of noise status.
Such information is generated every 15 minutes, based on the data-quality state vector, which is recorded as time-series data with bit information at a 16 Hz sampling rate.
KAGRA's segment information is sent to a database server called ``DQSEGDB'' \cite{DET:ref_DQSEGDB} at the California Institute of Technology (CIT)  and is stored together with segment information from LIGO, Virgo and GEO.
Some search pipelines use such segment information from multiple detectors to perform coincidence and coherence searches. These segments will actually be used in the offline searches of O3GK.
For future observing runs, we plan to create not only information indicating whether or not a segment is in science mode, 
but also information containing various noise conditions caused by earthquakes, loud microseismic disturbances, and so on.

\subsection{Transient-noise identification}\label{DET:ssec_glitch}
While gravitational-wave search pipelines usually use only the gravitational-wave channel data, the quality state vector, and segment information, other auxiliary channels help with noise investigations to reduce false candidate events caused by noise transients.
Especially for burst searches, in which theoretical waveforms are not assumed, removing false events by using the auxiliary signals is one of the most important tasks for the reliable detection of gravitational wave events.
Coincidence investigations of transient signals with the gravitational-wave channel and auxiliary channels are often performed to provide veto analysis for candidate transient gravitational wave events.
The method is called ``Hierarchichal Veto (hveto)'' \cite{DET:ref_hveto,DET:ref_hveto2}.
hveto rejects false events by using the significance of coincident noise events between the GW channel and auxiliary channels. 
In order to detect glitches in the GW channel and auxiliary channels, Omicron pipeline was used during O3GK for around 200 auxiliary channels.
The Omicron pipeline, based on the Q-transform method, detect transient events in a time series \cite{DET:ref_Omicron,DET:ref_Omicron2,DET:ref_QTF}.
This method provides GPS time when event occurred, central frequency, and Q-value for every transient event in the gravitational-wave channel and auxiliary channels.
The hveto analysis vetoes the gravitational wave candidate events caused by noise transients in auxiliary channels.

Veto analysis using hveto is being conducted as the offline analysis during O3GK.
An event list of noise transients was provided every 15 minutes as input to hveto.
For future observing runs, online veto analysis is also required.
\textcolor{black}{Data transfer for online searches 
including calibration, $h(t)$ generation, and the duration of data files 
takes less than one minute to the main data centers at Kashiwa and overseas sites such as CIT} \cite{KAGRA:PTEP04}.
Depending on the computing time in the search pipeline itself, the veto process can be started within a 10 - 20 minute delay, which is necessary to provide breaking news of GW-event alerts.
In the future, we also aim to reduce the time spent both in data transfer and on the GW search itself.
There are some plans to reduce the latency in the data transfer and the GW searches for the purpose of the multi-messenger astronomy. Shorter latency is required also in the investigation of the noise-transient and providing data quality information in order to provide the reliable GW alerts in future observation such as O4, O5 and so on. Actually KAGRA aim to provide the segment information with a few minutes cadence in O4 observation.

\section{Physical environmental monitors}\label{Sec:main:PEM}
\subsection{Introduction}\label{Sec:PEM:Introduction}

Because the typical amplitudes of GWs are extremely small, strains on the order of 10$^{-21}$, 
in principle, small vibration from instruments, small sound from outside of the experimental area and so on can produce noise-source contamination that reduces the sensitivity.
Major noise sources include environmental disturbances caused by earthquakes, effects from magnetic and acoustic fields,temperature fluctuations, and so on.
To evaluate the noise sources, about 100,000 auxiliary channels are recorded by
the KAGRA digital system.

The three main purposes of physical environmental monitorings are the following:
The first use of PEMs is to identify noise sources and understand their couplings to  
detector sensitivity so that noise-hunting measures can be applied\cite{PEM:LIGOS6}.
The second purpose is to collect environmental information
that can be used in evaluating the data quality of the GW channel
and in trying to distinguish GW signals from any pseudo signals 
caused by noise.
The details are described in Sec. \ref{DET:ssec_glitch}.
The third purpose is for R$\&$D studies directed toward 
the development of 3$^{rd}$-generation GW interferometers.
As described above, the KAGRA interferometer has two unique features: 
the underground site and cryogenic technology.
Both features will be essential for 3rd-generation detectors. 
Understanding the influences of these new technologies on GW detectors is attracting great attention 
from the international LIGO and Virgo collaborations.

\subsection{Installation protocols for the KAGRA PEM sensors}\label{Sec:PEM:PEMs}

\begin{figure}[htbp]
    \centering
    \includegraphics[scale=0.55]{ 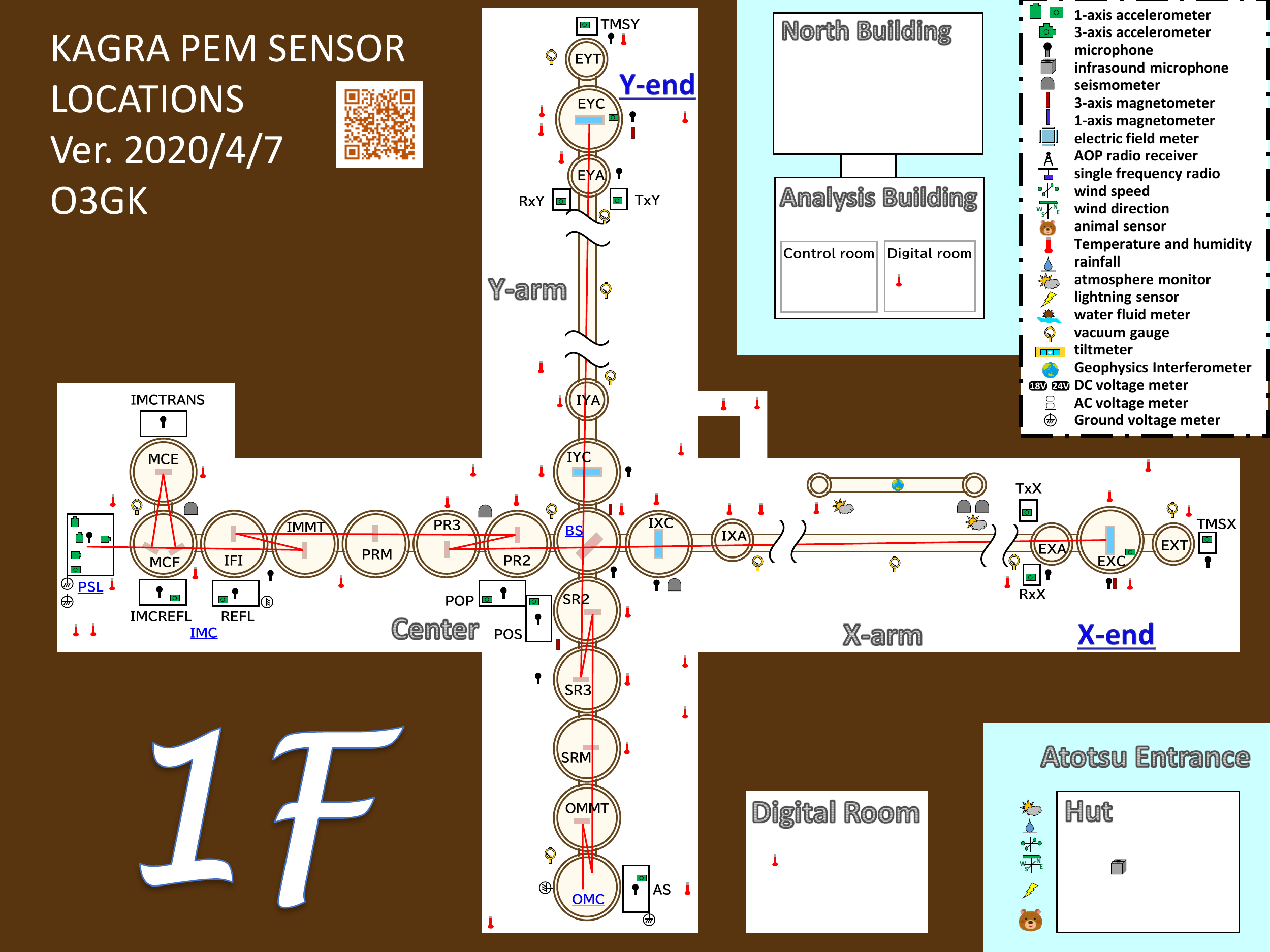}
    \caption{Location map of the KAGRA PEM sensors at first floor (1F) for the O3GK observation (the design is based on LIGO \cite{LIGO:PEM:S6}). This map is available at \cite{PEMmap}.}
    \label{Fig:PEM:PEMMAP}
\end{figure}

To evaluate the environmental noise, 
we have installed more than 100 PEM sensors 
in the KAGRA experimental site 
(including outside the tunnel). 
Detailed information about the sensors used for the O3GK observation is summarized in Table \ref{Tab:PEM:PEMs}, 
including the sensor type, product name, operating frequency, 
and number of sensors,
and in Fig. \ref{Fig:PEM:PEMMAP} with a location map.
Signals from the fast sensors (seismometers, accelerometers, microphones, magnetometer, and voltmeter) are acquired by the KAGRA digital system together 
with the interferometer signals and suspension signals.
The slow sensors (thermohygrometers and weather station) 
have their own data loggers, and the signals 
are also merged into the KAGRA data through the
EPICS\footnote{Experimental Physics and Industrial Control System, https://epics-controls.org} 
system. 

\begin{table}[htbp]
  \caption{Summary of the KAGRA PEM sensors installed for the O3GK observation.}\label{Tab:PEM:PEMs}
  \begin{tabular}{|c|c|c|c|} \hline
    Sensor type     & Product name               & Operating frequency  & Number  \\  \hline \hline
    Seismometer 1   & Trillimu120Q                  & 10 mHz -150 Hz    &  3 \\ \hline
    Seismometer 2   & Trillium compact              & 10 mHz -150 Hz    &  3 \\ \hline
    Accelerometer 1 & TEAC 710                      & 20 mHz - 200 Hz   & 10 \\ \hline
    Accelerometer 2 & TEAC 706                      & 3 Hz - 14 kHz     &  6 \\ \hline
    Accelerometer 3 & PCB M601A02                   & 17 mHz - 10 kHz   &  4 \\ \hline
    Accelerometer 4 & KISTLER 8640A5                & 0.5 Hz - 3 kHz    &  4 \\ \hline
    Microphone 1    & B$\&$K 4188-A-021             & 20 Hz - 12.5 kHz  &  3 \\ \hline
    Microphone 2    & ACO microphones               & 20(1) Hz - 20 kHz & 17 \\ \hline
    Microphone 3    & Audio-technica AT-VD6         & 60 Hz - 15 kHz    &  2 \\ \hline
    Magnetometer    & Bartington Mag-13MCL100       &  DC - 3 kHz       &  3 \\ \hline
    Voltmeter       & KAGRA ADC (directly)          &  DC - 16 kHz      &  5 \\ \hline
    Thermometer     & T$\&$D RTR-507SL              & 5 min sampling    & 77 \\ \hline
    Weather station & Davis Vantage Pro2 $\#$6152JP & 1 min sampling    &  1 \\ \hline
    Lightning sensor& Blitzortung System Blue       & (triggered time)  &  1 \\ \hline
  \end{tabular}
\end{table}

The PEM sensors are installed for the following protocols in KAGRA:

\subsubsection{Monitors for vibration, sound, and the voltage at the optical tables}\label{Sec:PEM:optlcaltable}
Monitoring and controlling the auxiliary optics is 
important for interferometer operation.
Such auxiliary optics are used for many purposes, such as 
laser-source stabilization, optical mode matching, 
sensing for the interferometer controls, 
and control of the photon calibrator. 
For sensing and stabilization, 
multiple optical tables are installed in several places.
Optical parts like photo-diodes and periscopes are fixed onto the table,  
so they are directly affected by their environments.
We placed at least one accelerometer, microphone, and voltage monitor
to monitor the electrical ground between each optical table and its ADC.
By monitoring those signals, we can identify stationary interferometer noise, 
narrow band frequency noise (line noise) and glitch noise caused by the environment,
such as the acoustic noise shown in Fig. \ref{Fig:PEM:injection}

\subsubsection{Monitors for ground motions in the underground facility}\label{Sec:PEM:ground}

We placed three tri-axial seismometers at the center, X-end, and Y-end areas  
and positioned one GIF along the X-arm (the GIF details are described in Section. \ref{Sec:main:GIF})
to monitor the ground motions caused by Earth tides, earthquakes, ocean waves, and human activities.
An important point is that the seismometers are placed on the 2nd floor of each area; 
the four cryogenic mirrors that comprise the Fabry-Perot cavities,  
are hung from the 2nd floor.
They are used not only ground motions but also for sensor corrections\cite{FujiiThesis}, 
controlling the suspensions with multiple sensors.

In addition, we installed three compact seismometers on the 1st floor of the center area: 
(1) near the input-mode cleaner (IMC), 
to monitor local ground motions at the pre-stablized laser (PSL) room, the IMC, 
and the input mode-matching telescope, 
(2) near the beam splitter (BS), to monitor local ground motions at the power-recycled mirror chambers, the BS chambers, and the signal-recycled mirror chambers, and
(3) near the Input test mass Chamber X (IXC), to monitor local ground motions at the
cryostat and 
to compare differences between the 1st and 2nd floors.

\subsubsection{Monitors for magnetic fields in the underground facility}\label{Sec:PEM:magnetic}

Magnetic-field noise is an important environmental noise for a GW detector, 
because it can cause electrical noise due to mirror motions.
At LIGO and Virgo, the identification and mitigation of narrow spectral artifacts 
--due to power lines and magnetic fields to/from suspensions or electrical circuits--
played important roles in O1 and O2 \cite{LIGO:mag}.
It is even more important for KAGRA to monitor the magnetic fields in the experimental site, because coil-magnet actuators are used to control the suspensions 
instead of the electro-static drivers used test mass in LIGO\cite{LIGO:control}.

We placed three 3-axis magnetometers near the BS chamber, X-end cryo-chamber, and Y-end cryo-chamber
to monitor the magnetic fields coming from various instruments (\textit{e.g,} cryo-coolers, power lines, and digital devices) or due to natural phenomena (\textit{e.g,} lightning strokes, magnetic storms, and Schumann resonances \cite{schumann}).

\subsubsection{Monitors for room temperature and humidity in the underground facility}\label{Sec:PEM:temp}
Even though the temperature of the underground site is stable 
compared with outside,
the KAGRA suspensions are extremely sensitive to the surrounding temperature.
Because many delicate analog circuits are used, 
monitoring the humidity is also important.
The temperature and humidity vary as instruments are turned on and off (\textit{e.g,} vacuum pumps and fans).
We placed a number of thermo-hygrometers on all the electrical racks, in the clean booths, near the chambers, and near the air conditioners \cite{ondotori}.

\subsubsection{Monitors for weather conditions outside the tunnel}\label{Sec:PEM:weather}
It is known that the weather condition is correlated with detector noise, 
for example, wing speed, Barometric pressure.
To monitor the environment outside the KAGRA tunnel, a weather station 
was set up in front of the tunnel entrance. 
It monitors the temperature, humidity, air pressure, rainfall, wind speed, and wind direction. 
In addition, a lightning sensor was installed as a part of the Blitzortung.org network \cite{lightning}
to monitor the time and position of lightning strikes.

\subsection{Development of a portable PEM system}\label{Sec:PEM:portable}

\begin{figure}[htbp]
    \centering
    \includegraphics[scale=0.9]{ 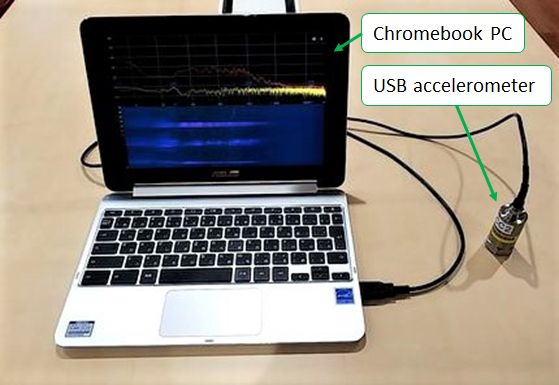}
    \caption{Photograph of a Chromebook with a USB accelerometer. The real-time spectrum and a spectrogram generated by free software are displayed.
    One of the most strong point of this system is its compactness. We can move this system easily.}\label{Fig:PEM:Chromebook}
\end{figure}

Besides the regular PEM sensors at various fixed locations, 
we are also utilizing so-called `Portable PEMs' 
in addition to the regular PEM sensors
to assess various unknown noise sources, 
to make characterization of the KAGRA instruments easier, and 
to understand the noise-coupling paths.
There are two types of portable PEMs:
One is a combination of an analog output sensor and the KAGRA digital system, as with the regular PEM sensors. 
Some versatile ADC channels are reserved for this purpose in each area. 
Since the digital system is available, it is possible to carry out data analyses 
with other channels; e.g, to provide real-time coherent analyses.
The other one is a combination of a USB sensor and a Chromebook\textregistered \ PC (ASUS Flip C101PA), 
as shown in Fig. \ref{Fig:PEM:Chromebook}.
Since this PC has USB-A and USB-C ports, and since Android\textregistered \ applications are available, 
a real-time spectrogram from a USB sensor (microphone, accelerometer, and magnetometer) can be displayed.
Using this system, we can work free from any limitation due to cabling, power supply, ADC port, etc. 
This system enables us to investigate environmental noise very effectively.
Detailed information about the portable PEM system will be described in a future paper.

\subsection{PEM injection}\label{Sec:PEM:PEMAnalysis}
PEM injection is an important measurements for evaluating 
environmental effects, 
--such as sound and magnetic fields, 
vibration from instruments, and RF signals--
on the detector sensitivity.
The coupling function $C(f)$ is given by
\begin{equation}
C(f) = \sqrt{ \frac{\tilde{Y}_{\mbox{inj}}^{2}(f) - \tilde{Y}^2(f)}{\tilde{X}_{\mbox{inj}}^{2}(f) - \tilde{X}^2(f)}}, 
\end{equation}
where $\tilde{Y}_{\mbox{inj}}(f)$ and $\tilde{Y}(f)$ are the amplitude spectral density of the GW strain channel
with and without PEM injections, respectively, and 
$\tilde{X}_{\mbox{inj}}(f)$ and $\tilde{X}(f)$ are the amplitude spectral density of the PEM sensor signal. 
The effect of environmental noise on the sensitivity is given by
\begin{equation}
Y_\mathrm{PEM}(f) = C(f)\cdot \tilde{X}(f) 
= \sqrt{ \frac{\tilde{Y}_{\mbox{inj}}^{2}(f) - \tilde{Y}^2(f)}{\tilde{X}_{\mbox{inj}}^{2}(f) - \tilde{X}^2(f)}}  \cdot \tilde{X}(f).
\end{equation}
These formulas are also used by LIGO\cite{PEM:LIGO} and Virgo\cite{PEM:Virgo}.

\begin{figure}[htbp]
    \centering
    \includegraphics[scale=1]{ 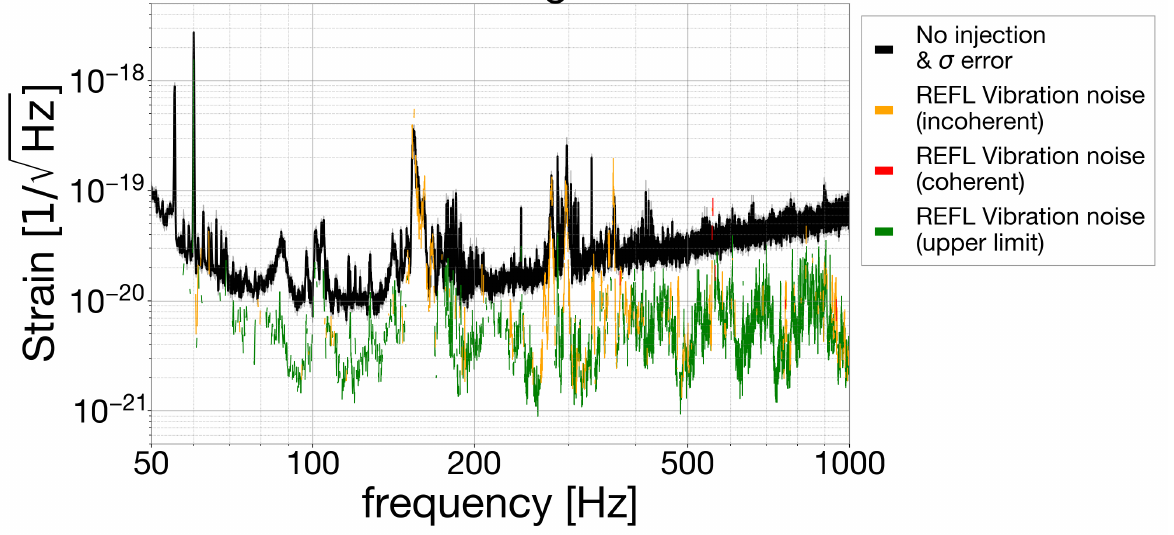}
    \caption{Result of an acoustic-injection test performed during FPMI commissioning in December 2019 \cite{TTThesis}. 
    Strain sensitivity without PEM injection (black), with projected acoustic noise (orange, incoherent; red, coherent), and the $3\sigma$ upper limit to the acoustic noise (green).
    The several peaks in this figure can be identified with acoustic noise sources.
    }\label{Fig:PEM:injection}
\end{figure}

Figure \ref{Fig:PEM:injection} shows the results of an acoustic-injection test performed during FPMI commissioning in December 2019 
as one example of a PEM injection into KAGRA \cite{TTThesis}. 
The several peaks in this figure can be identified with acoustic noise sources
around the optical tables.
More detailed studies with the PRFPMI configuration were performed before and after the O3GK observation, and a paper describing the results is in preparation.

\subsection{Noise hunting using PEMs} \label{Sec:PEM:NoisePEM}
We succeeded in hunting down several noise sources 
that affected the interferometer sensitivity.
Representative noise-hunting are summarized below.

\begin{itemize}

\item 17.2 Hz noise hunting using installed PEM sensors\\
Noise was detected at 17.2 Hz in the interferometer control signal.
It was largely coherent with the optical levers that monitor the motions of the test masses, at the signal-recycling mirrors. 
We found that the fan filter unit (FFU), which is used to keep the clean booth at 
a given clean level, generated the vibration. 
The resonant frequency of the framework of the clean booth 
turned out to be 17.2 Hz.
When the FFU was turned off, the noise vanished.

\item 44 Hz noise hunting by portable PEM system\\
Noise was detected at 44 Hz in the frequency noise of the 
auxiliary lasers that support arm-length stabilization for interferometer-lock acquisition  \cite{KAGRA:greenlaser}. 
When we evaluated the coherence with the power of those
auxiliary lasers using the PEMs, 
we found that the accelerometers placed in the PSL room exhibited large coherence.
Using a portable PEM with Chromebook, the large vibration at 44 Hz was
identified to be the mechanical vibration of a 24V
DC power supply used for the laser shutter.
We changed the position of the power supply, and this noise disappeared.

\item 160, 280 and 360 Hz noise hunting using PEM injection \\
The bumps around 160, 280, and 360 Hz in Fig\ref{Fig:PEM:injection} 
were identified as ambient acoustic noise in the FPMI configuration; 
similar results were observed in the PRFPMI configuration before the O3GK run. 
By using a hammering test we found that they came from the bellows at the IMC output 
(most likely scattered-light noise).
We suppressed these noise sources by reducing the sound sources and adding sound proofing.

\end{itemize}

\section{Geophysics interferometer (GIF)}\label{Sec:main:GIF}
\subsection{Introduction}\label{sec:GIF:introduction}
\begin{figure}
\begin{center}
\includegraphics[width=8cm]{ 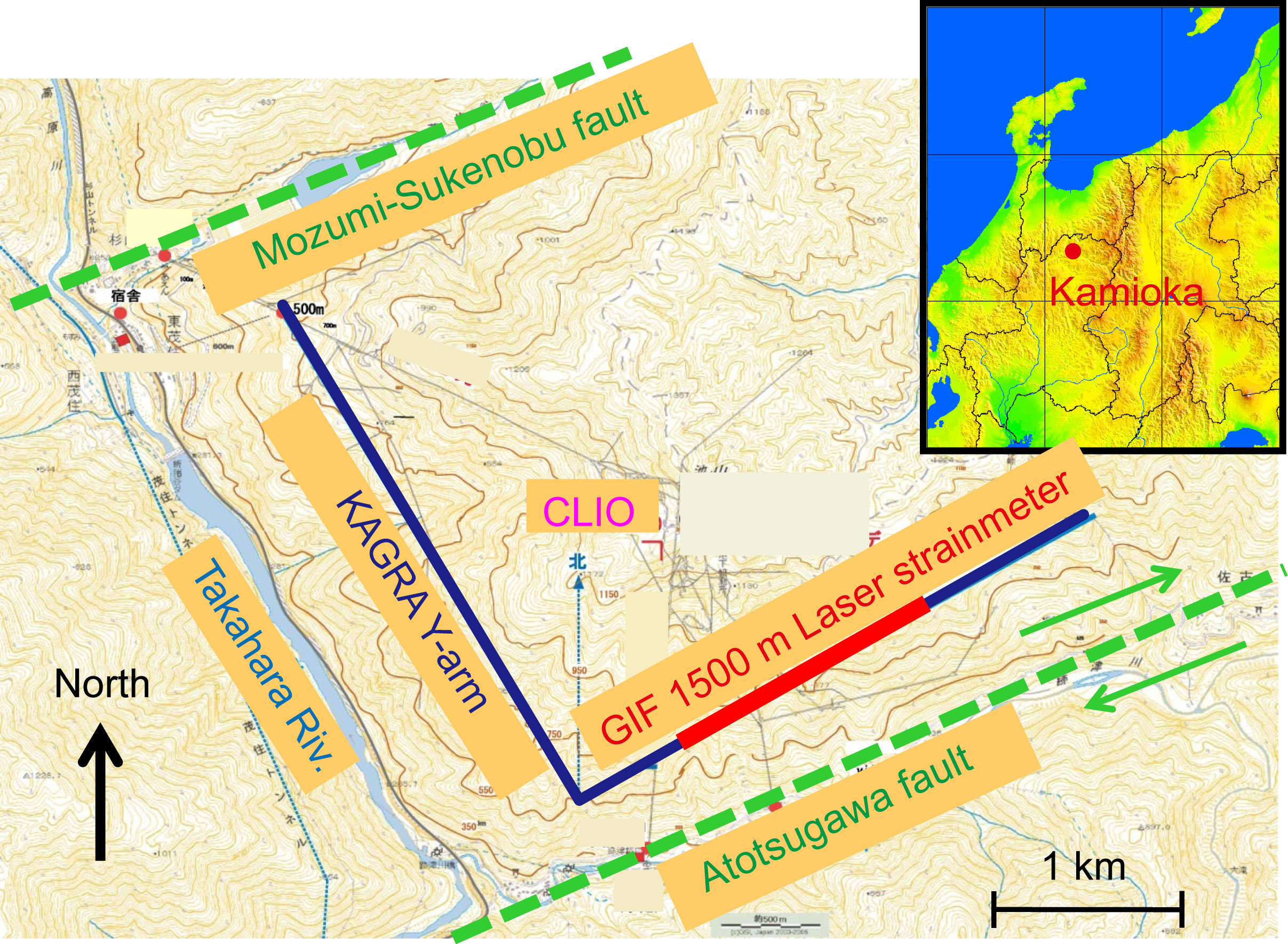}
\caption{Location of the GIF 1500-m laser strainmeter in the KAGRA X-arm tunnel and surrounding area. Adopted from reference \cite{GIF:GIF2017}}
\label{fig:GIF_location}
\end{center}
\end{figure}

  The GIF is one of KAGRA's unique features. It is a pair of Michelson laser interferometers specifically designed to measure ground motions (strains) along the KAGRA arms. The GIF covers a wide frequency range, which includes effects such as tidal motions, microseismic motions, coseismic steps, Earth’s free oscillations, slow earthquakes, and so on\cite{GIF:Araya2010}\cite{GIF:Benioff}\cite{GIF:Dragert}. These events themselves are of interest for geophysical studies, and in addition, the ground motions detected by the GIF can be used to compensate for changes in the KAGRA  baseline lengths in order to improve its stability. The first GIF strainmeter was constructed in the KAGRA X-arm tunnel in 2016, and it has been in operation since then. See Fig.\ref{fig:GIF_location} for its location.

\subsection{The GIF system}\label{sec:GIF:description}
\begin{figure}
\begin{center}
\includegraphics[width=\hsize]{ 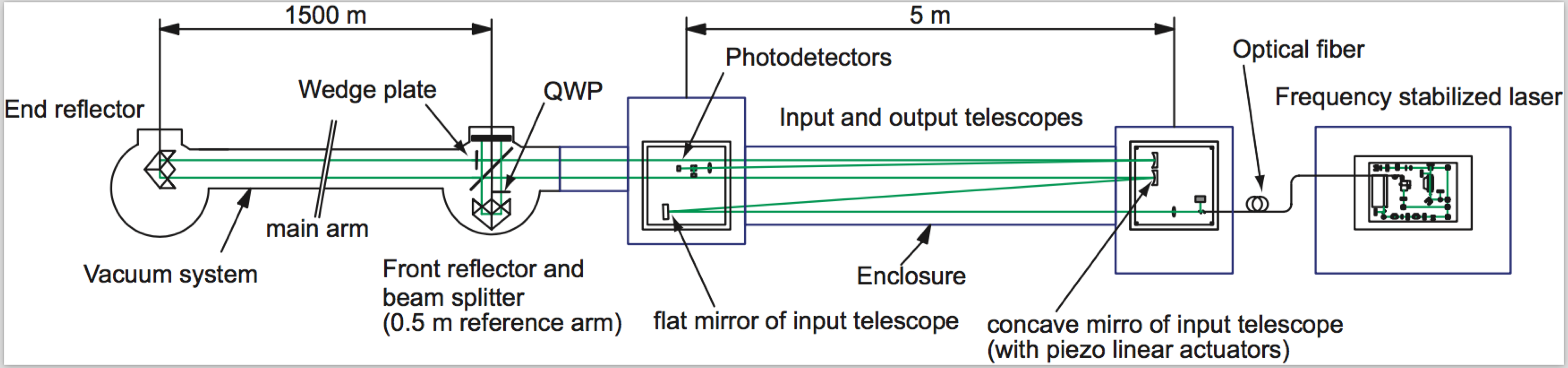}
\caption{Optical configuration of the GIF. The interferometer arms are located in vacuum. The input and output telescopes are placed in the atmosphere but are covered by hard enclosures.}
\label{fig:GIF_layout}
\end{center}
\end{figure}

 The basic design of the GIF strainmeter is an asymmetric Michelson interferometer with 1.5 km- and 0.5 m-long arms. The interferometer optics consist of two retroreflectors, a BS, a quarter-wave plate (QWP), and a wedge plate, as shown in Fig.\ref{fig:GIF_layout}. The optics are housed in vacuum chambers located at both ends of the main arm. They are separated by 1.5 km and are connected by a vacuum tube. The vacuum pressure in the optical path is maintained lower than $10^{-2}$ Pa to suppress optical-path-length fluctuations due to variations in the refractive index of the residual gas. The 0.5 m-long reference arm consists of the BS and one of the retroreflectors, both mounted on a single Super Invar plate for thermal stability. The other reflector is installed in another vacuum chamber, and together with the BS, it forms the main arm, the displacement of which is detected by the interferometer. The optical components are rigidly connected to bedrock, and since no length control is applied to them, they follow the exact ground displacements.
 
 A frequency-doubled Nd:YAG laser is used as the light source. The laser frequency is stabilized to an absorption line of iodine ($\rm{I_2}$) gas via the saturated-absorption technique. Frequency fluctuations directly cause displacement noise due to the asymmetry of the interferometer \cite{GIF:I2laser}. The fundamental limit to the strain resolution of this instrument is set by the stability of the laser frequency. The actual frequency-noise level is estimated to be better than $10^{-11}$ over a 10 second period by comparison with an identical stabilized laser. The resolution is sufficient to observe the ground motions at low frequencies which are dominated by the tides and microseisms, in order to provide baseline compensation for KAGRA.

 The laser beam is introduced into the input optical system through a polarization-maintaining fiber. The input optics consists of a pair of lenses and a flat and a concave mirror, which form a folded mode-matching telescope. This arrangement optimizes the beam profile so that the beam waist is located at the end reflector, and the return beams from the two arms overlap adequately on the BS. The diameters of the beam waist and the return beam from the main arm (on the BS) were calculated to be 32 mm and 45 mm respectively \cite{GIF:MiyoThesis}. The visibility of the interferometer is maintained by aligning the input beam with the main arm by tilting the concave mirror with piezo linear actuators (Picomotors, Newport Inc.). This realignment procedure is regularly (typically once per month) performed from a remote laboratory in Tokyo over the Internet. A similar optical system is installed along the input telescope to form an output telescope that focuses the return beam onto the photodetectors (PDs). The input and output optics are mounted on two optical tables separated by 5 m, and the optical path between them is doubly covered by PVC pipes and by an enclosure with aluminum-plate walls in order to prevent contamination and airflow that causes alignment fluctuations.

A quadrature-detection technique is used to obtain the phase changes of the interferometric fringes that represent the ground displacements, including their directions. Combined with the absence of length control, this configuration enables a very wide (ideally infinite) length-observation range. The QWP inserted in the reference arm makes this technique possible by creating two linearly-polarized components that are 90 degrees out of phase, and they are detected by two PDs at the output port after being separated by a polarized beam splitter.

We developed a data acquisition (DAQ) and automatic control system for laser stabilization based on a commercial modular controller (PXI system, National Instruments Inc.). It records the interferometer signals, i.e., the fringe signals and other monitoring signals (50k samples per second), together with environmental-monitoring signals (200 samples per second). The controller sets the status of the laser-frequency stabilization system, which is implemented with analog circuits, in the lock-acquisition or lock-maintaining mode to achieve a duty cycle of 99.4 \% (average in 2019).

\subsection{Details of implementation, installation, and operation}\label{sec:GIF:implementation}

 The GIF is constructed in the KAGRA tunnel in a severe environment, with water dripping frequently from bare rock surfaces, and the atmosphere is very humid and dusty. In order to protect the interferometer optics and the laser system from contamination, we built clean booths with clear PVC walls around the vacuum chambers and the optical tables prior to their installations.
 
\begin{figure}
\begin{center}
\includegraphics[width=10cm]{ 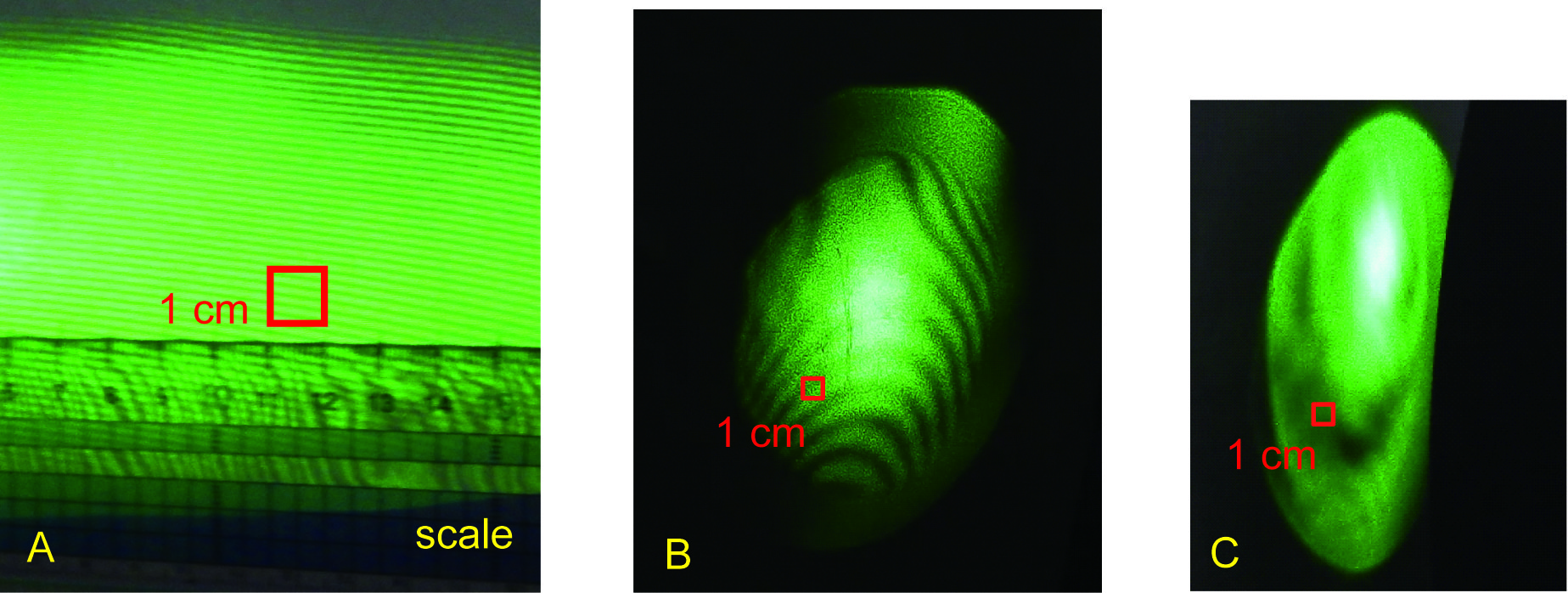}
\caption{Progress of wavefront correction. (A) Five or six fringe stripes/cm were observed without correction, which resulted in insufficient visibility to obtain phase information. (B) That number was reduced to 1 stripe/cm by adjusting the angle of the BS plate, enabling phase determination. (C) Further correction was achieved by inserting a wedge plate in the main arm to improve the visibility.}
\label{fig:GIF:wavefront}
\end{center}
\end{figure}

\begin{figure}
\begin{center}
\includegraphics[width=7.5cm]{ 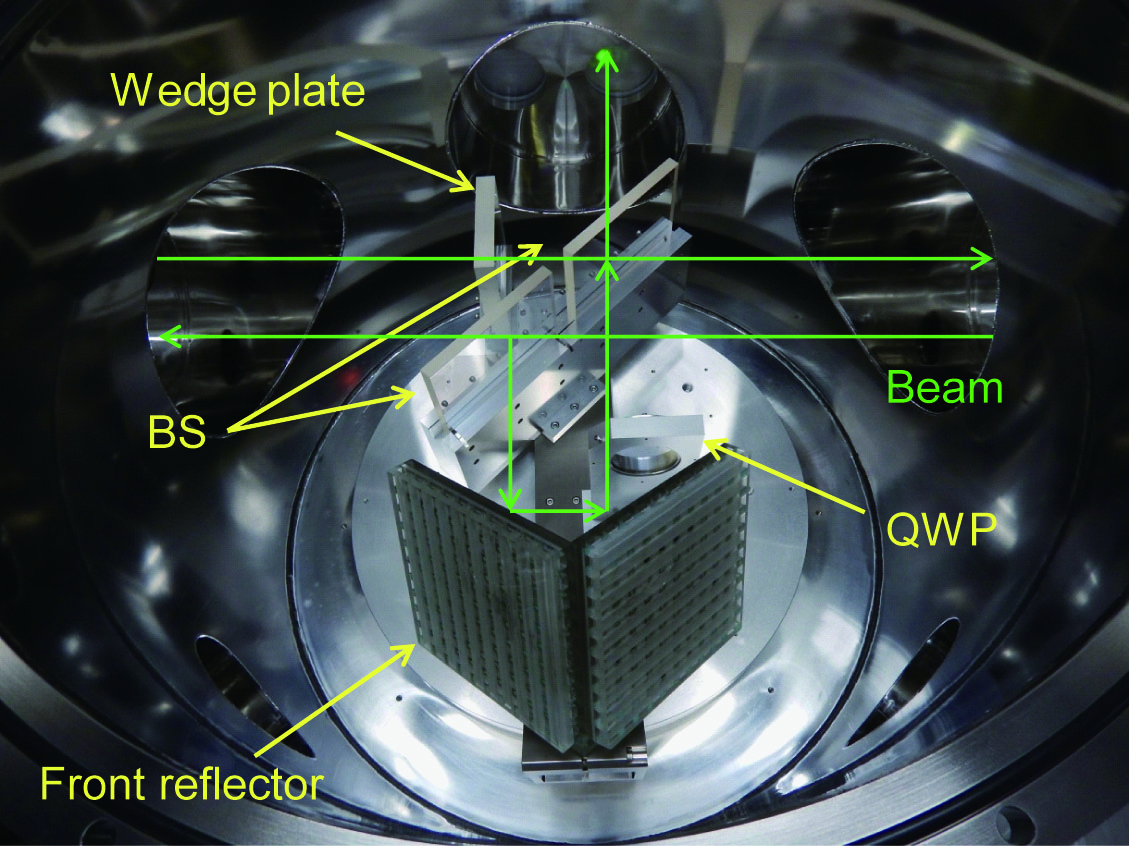}
\caption{Inside the front vacuum chamber. The BS and retroreflector are mounted on a Super Invar platform to form a 0.5 m reference arm. The wedge plate provides wavefront correction.}
\label{fig:GIF:fronttank}
\end{center}
\end{figure}

 The long baseline length of the GIF is advantageous for achieving better strain resolution, but due to beam divergence it requires larger optics than shorter interferometers. This introduced difficulties in the production of some optical components, such as the retroreflectors (which require 15-inch clear apertures) and the BS. The  parallelism and flatness of their surfaces strongly affect the fringe visibility of the interferometer. For the retroreflectors, we made a simple two-dimensional model to determine the requirements for surface flatness necessary to realize the desired fringe visibility. However, technical limitations in their production prevented us from meeting these requirements fully, so the reflectors were made using best efforts. After their production, we recalculated the expected visibility to be 0.53, based on the surface-flatness distribution measured by the manufacturer. Due to additional degradation imposed by other components, the actual visibility was reduced to 0.1, but this is still sufficient to extract the necessary phase information. A similar problem occurred in manufacturing the BS. In the initial design we had planned to make it from a single glass plate, expecting better parallelism, which is important for achieving a uniform wavefront (i.e., better visibility). However, it turned out that a single plate large enough to cover both the input and output beams was too large for the fabrication facility of the manufacturer. We therefore decided instead to make two separate plates, one each for the input and output beams. This ``compromise'' actually allows us to adjust their angles independently by inserting thin spacers into their mounts, coarsely correcting the wavefront distortion of the returning beams from the main and reference arms (Fig. \ref{fig:GIF:wavefront}). Additional wavefront correction was applied by inserting a glass wedge plate between the BS and the main arm reflector (Fig. \ref{fig:GIF:fronttank}) to compensate for residual wavefront mismatch.
 
 The lock status of the laser-frequency stabilization is continuously monitored by the DAQ system. In order to maximize the observation time, it starts the relocking process immediately after a loss of lock is detected. Due to the automatic relocking system and the stable environment of the underground site, the GIF requires little human effort to maintain its operation. We use monthly realignment of the input beam to compensate for its drift in tilt (supposedly caused by plastic deformations of the  springs used in the flat mirror mount of the input telescope). The beam path in the saturated-absorption optics needs realignment only a few times a year. These realignments can be done remotely without disturbing the site environment. We regularly check the status of the vacuum system, inspect the facility, and fix problems -- for instance, by installing shields for the vacuum components to protect them from water drops -- in order to maintain stable operation.

\subsection{Recent topics}\label{sec:GIF:recent}
\subsubsection{A study of barometric effects}\label{sec:GIF:baro}

\begin{figure}
\begin{center}
\includegraphics[width=8cm]{ 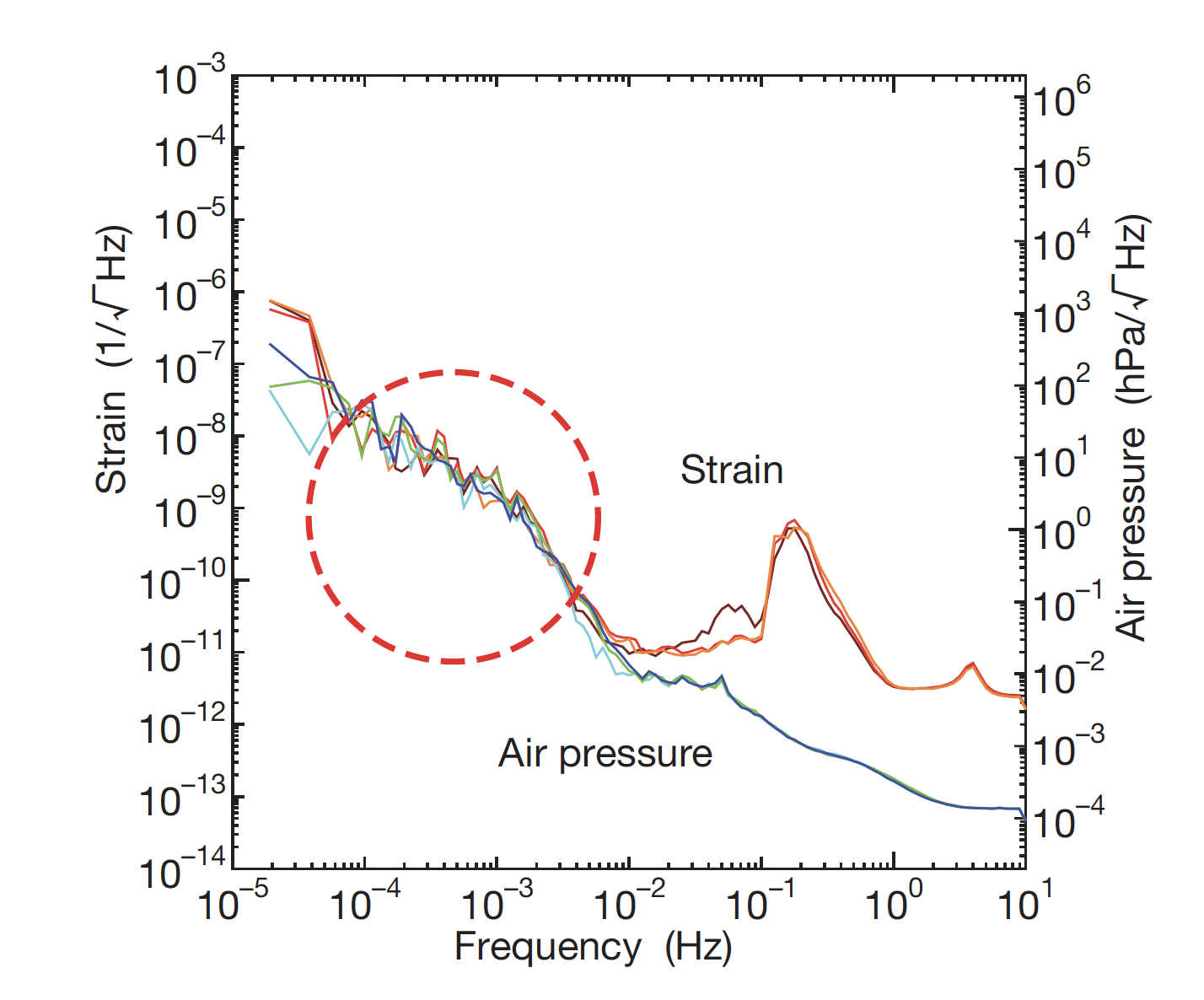}
\caption{Spectra of ground strains observed by the GIF and of local air pressure measured at the front and end chambers of the GIF. Strains in the 10$^{-4}$ - 10$^{-3}$ Hz region (within the dashed red circle) have a spectral shape similar to the air pressure. The barometric coupling to strain noise is estimated to be $\sim$ 0.55 $\times$ 10$^{-9}$ / hPa. Different datasets are shown in different colors to see the repeatability and the fluctuation.}
\label{fig:GIF:baro1}
\end{center}
\end{figure}

 Ground strain measurements at low frequencies are often influenced by variations in the air pressure \cite{GIF:baro}. Figure \ref{fig:GIF:baro1} shows the spectra of ground strains observed by the GIF and of the local air pressure measured at the front and end chambers of the GIF. Strains in 10$^{-4}$ - 10$^{-3}$ Hz region have a spectral shape similar to that of the air pressure, and their temporal variations also are highly correlated with the temporal variations in air pressure.
 
\begin{figure}
\begin{center}
\includegraphics[width=8cm]{ 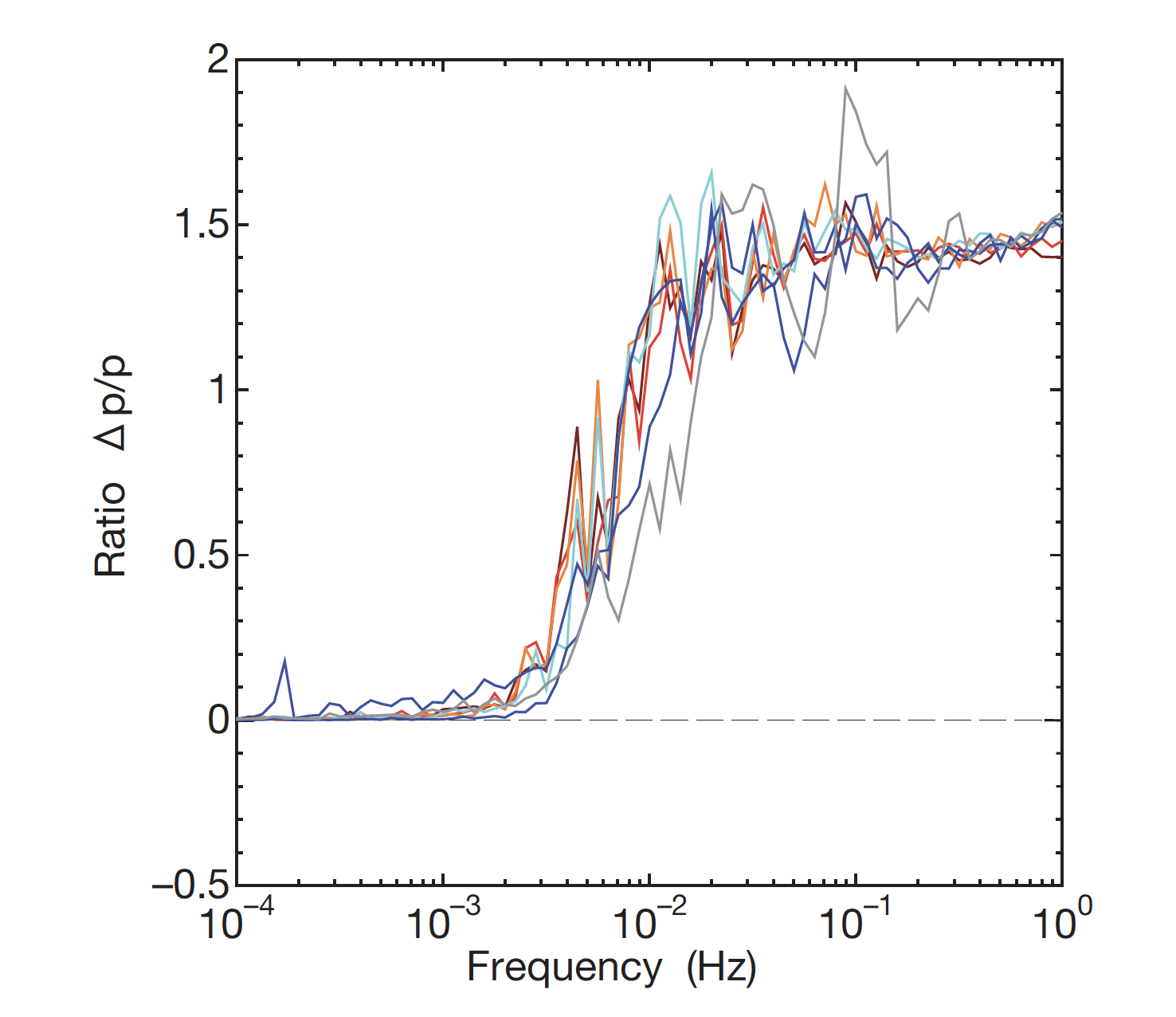}
\caption{Relative differences in air pressure between the front and end chambers of GIF. Both air pressures are almost identical (within 10 \% difference) below $\sim$ 3 mHz and they are uncorrelated above $\sim$ 10 mHz. Different datasets are shown in different colors to see the repeatability and the fluctuation.}
\label{fig:GIF:baro2}
\end{center}
\end{figure}

\begin{figure}
\begin{center}
\includegraphics[width=8cm]{ 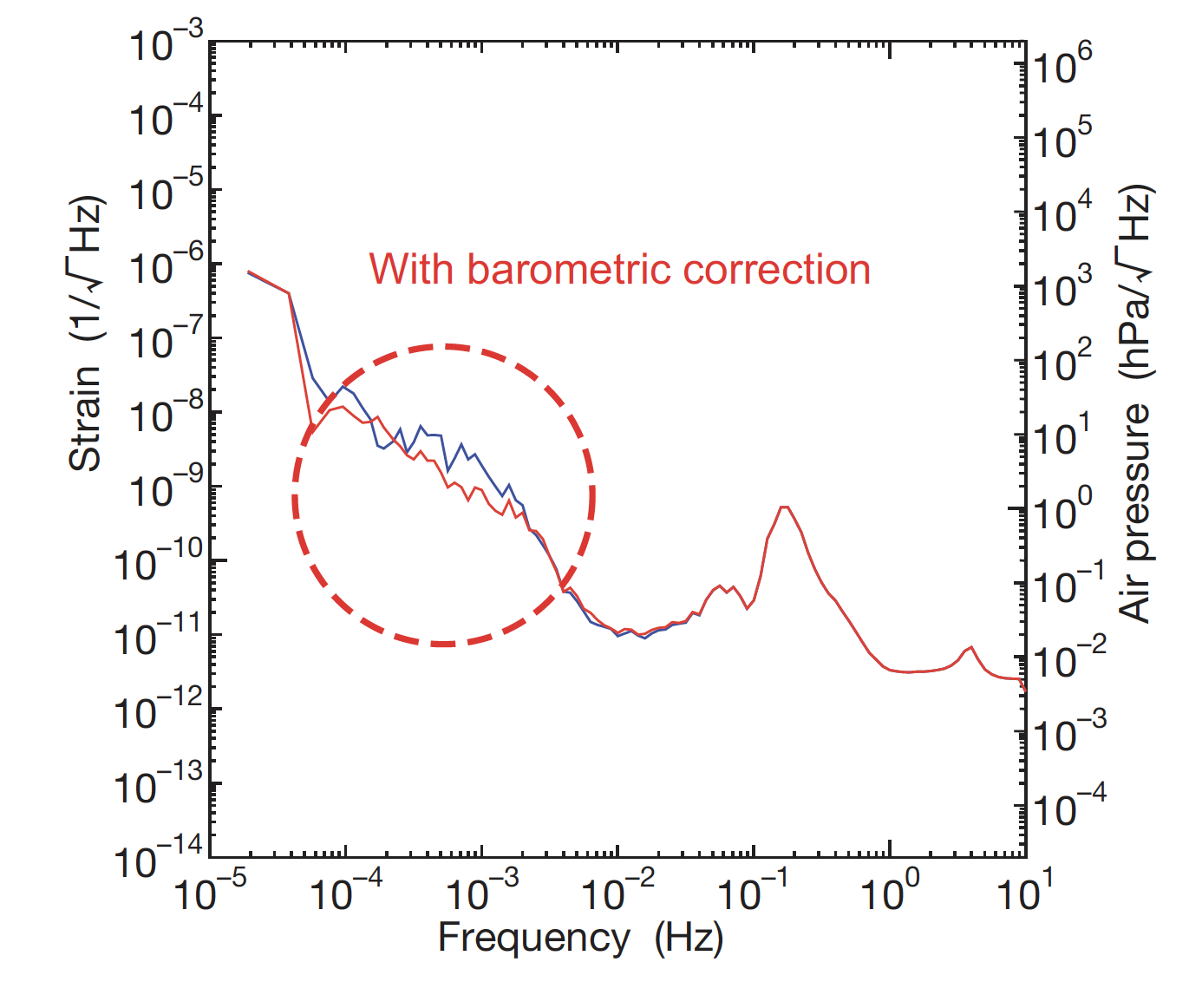}
\caption{Ground-strain spectra  observed by the GIF before (blue) and after (red) correction using the measured air pressure. The background strain is reduced by $\sim 1/3$ at best in the in 10$^{-4}$ - 10$^{-3}$ Hz region (within the dashed red circle). This limited reduction suggests that the background strain is not determined simply by the local air pressure but also is affected by the regional air pressures which may have different correlations to the local air pressure.}
\label{fig:GIF:baro3}
\end{center}
\end{figure}

 The barometric coupling to strain noise, in terms of a coefficient of strain response to air pressure, is estimated to be $\sim$ 0.55 $\times$ 10$^{-9}$ / hPa, which is consistent with typical values \cite{GIF:baro}. Air pressures at the front and the end chamber, which are separated by 1.5 km in the tunnel, are almost identical (within 10 \% difference) below $\sim$ 3 mHz, while they are uncorrelated above $\sim$ 10 mHz (Fig. \ref{fig:GIF:baro2}). Correcting the ground strain with the measured air pressure in the 10$^{-4}$ - 10$^{-3}$ Hz region, however, reduces the background strain only by $\sim 1/3$ at best (Fig. \ref{fig:GIF:baro3}). It should be noted that the reduction is still limited even in the period of bad weather when amplitudes of the background strain increase in proportion to air pressure. Therefore, it is suggested that the background strain is not determined simply by the local air pressure but also is affected by the regional air pressures which will have similar amplitudes but may have different correlations to the local air pressure. Baseline corrections of the GW detector based on in-situ measurements of ground strains are effective, especially in the 10$^{-4}$ - 10$^{-3}$ Hz region (see the following section), where local measurements of air-pressure data and seismometer data are insufficient due to limitations in the spatial distributions and instrumental noise, respectively.

\subsubsection{Baseline-length compensation in KAGRA}\label{sec:GIF:baseline}

 The duty cycle of a GW detector is usually limited by seismic noise below 1 Hz, produced by earthquakes, microseisms, tidal motions, etc. \cite{GIF:LIGOquake}. Active vibration isolation systems based on seismometers have been used to improve the detector's duty cycle by suppressing the effect of those noise sources \cite{GIF:LIGOactive}. Detection limit of the seismometers at low frequencies can be partly mitigated by using tilt components from dedicated tiltmeters \cite{GIF:Venkateswara}. However, seismometers and tiltmeters have a fundamental problem, that is they cannot distinguish horizontal acceleration from gravity acceleration introduced by ground tilt; it limits the performance of active isolation systems in the low-frequency range. Baseline-length compensation using a strainmeter can avoid the problem.
 
  Tidal effects can be removed by applying a global tide model \cite{GIF:LIGOtide}, but other noise sources are intrinsically unpredictable, such as the air-pressure effect described in the previous section. Therefore it is crucial to use the actual ground motions observed at the GW detector site in order to build an effective baseline-compensation system. The degradation of the duty cycle due to low-frequency seismic noise can be mitigated by implementing a compensation system using the GIF, a sensor that can measure the actual change in baseline length with sufficient sensitivity all the way down to DC. We have demonstrated such a baseline length compensation system for the KAGRA X-arm, using the strain signal measured by the GIF in October 2019 \cite{GIF:MiyoThesis}. In our control system, the change in the baseline length was measured accurately by the GIF, and that signal was fed forward to the actuators installed at the suspension point of the end test mass in order to suppress the change in the arm length of the cavity.
 
\begin{figure}
\begin{center}
\includegraphics[width=8cm]{ 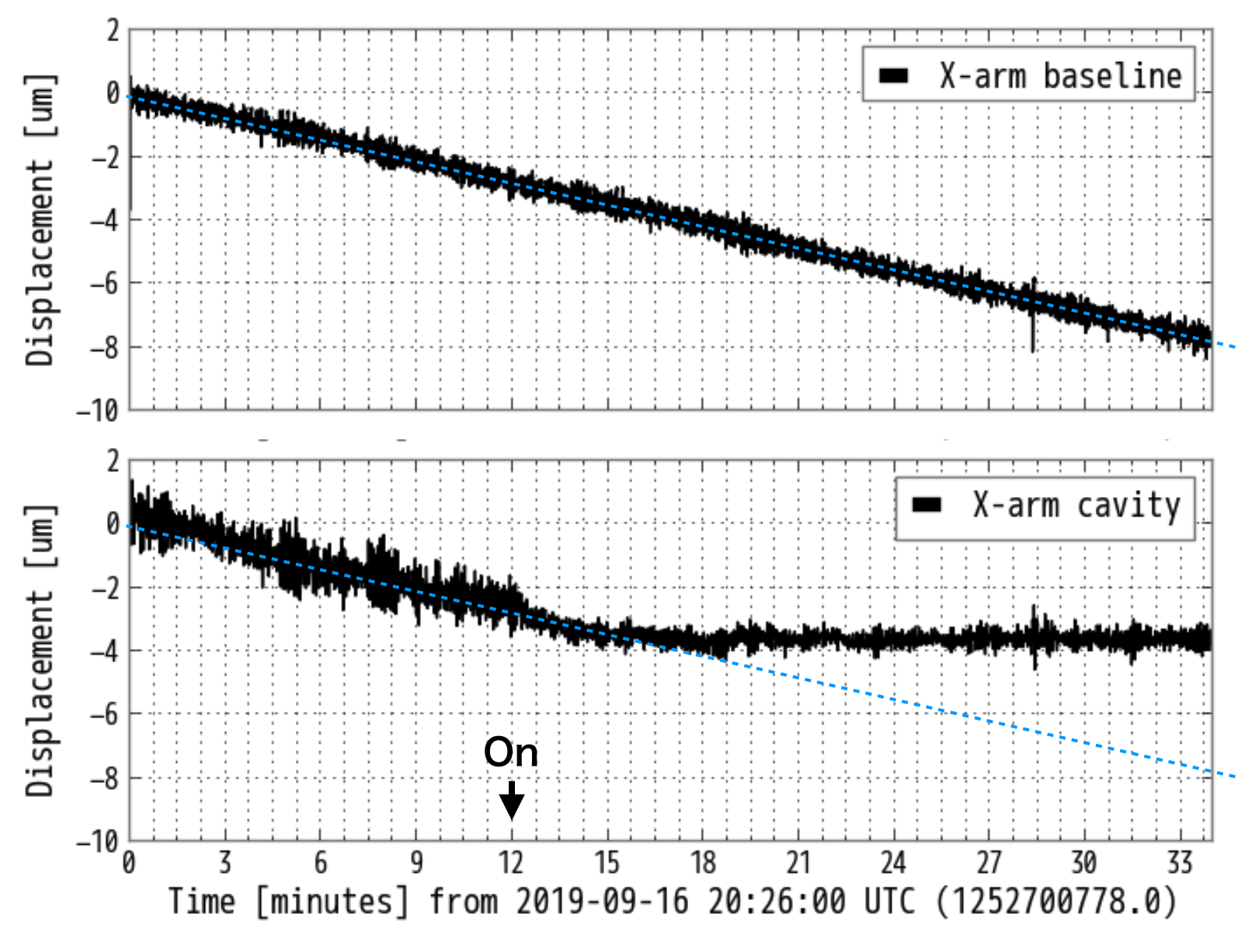}
\caption{Baseline motions observed by the GIF (top) and the change in length of the KAGRA X-arm cavity (bottom). Baseline-length compensation was turned on at 12 minutes (indicated by the ‘On’ arrow).}
\label{fig:GIF:baseline1}
\end{center}
\end{figure}

\begin{figure}
\begin{center}
\includegraphics[width=8cm]{ 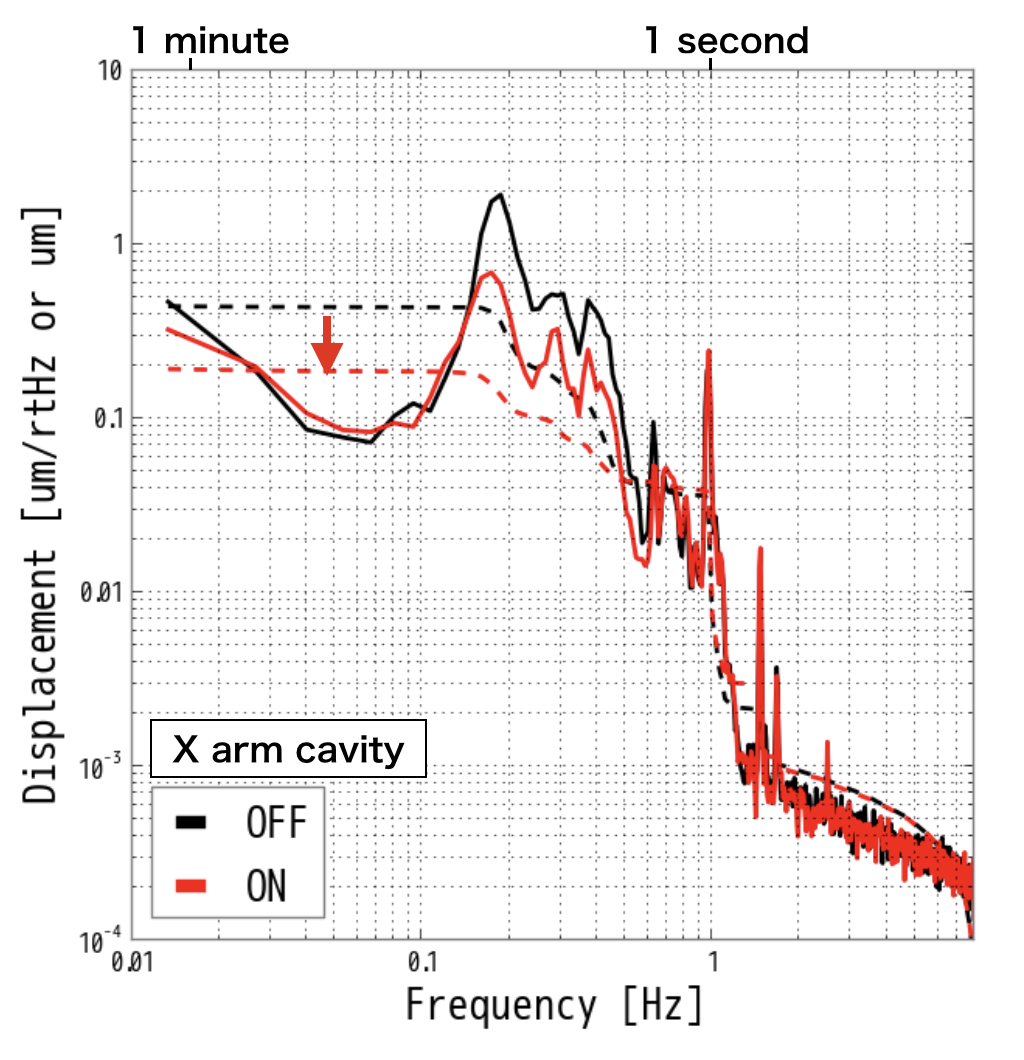}
\caption{Spectra of cavity-length changes of the KAGRA X-arm, before and after applying baseline-length compensation, black and red solid lines, respectively. The RMS motion (dashed lines in corresponding colors) was reduced by factor of $\sim$ 2.}
\label{fig:GIF:baseline2}
\end{center}
\end{figure}

 Figure \ref{fig:GIF:baseline1} shows the change in baseline length observed by the GIF (top) and the length change of the arm cavity (bottom). The constant drift in the top window corresponds to tidal motion at that time. The arm cavity was locked in resonance by controlling the laser frequency without applying any mechanical control, and the change in cavity length was derived from the control signal to the laser. Length compensation was turned on at the point indicated by ‘On’ in the bottom panel. There are two noticeable effects in the cavity-length signal after the control was introduced. The first effect is that the length change caused by the tidal motion was reduced to a few $\mu \rm{m}$. At least a one-order-of-magnitude reduction is estimated by comparing this number to the typical amplitude of tidal motion (several tens of $\mu \rm{m_{RMS}}$). The second effect is about a 50 \% suppression in the amplitude of the higher-frequency fluctuations. This was further studied in the frequency domain, and the spectra of cavity-length changes together with their RMS amplitudes with and without compensation are plotted in Fig. \ref{fig:GIF:baseline2}. The RMS amplitude was dominated by a microseismic peak around 200 mHz, and it was halved by the feedforward control.

\subsection{Summary}\label{sec:GIF:summary}
 The GIF strainmeter was designed to monitor ground motions over the 1.5 km baseline in the KAGRA tunnel, 
 and it has been operating with a high duty cycle. The strainmeter constantly observes tidal and microseismic motions and other occasional events, including near and far earthquakes as well as small coseismic steps originating from distant earthquakes. The strain resolution of the GIF was estimated to be  better than $10^{-12}$ in the 2 - 20 mHz range and $10^{-11}$ in the 1 mHz - 10 Hz range,
 based on the observed background noise (the lowest value among other laser strainmeters), which corresponds to ambient seismic motions or to laser-frequency noise, depending on frequencies \cite{GIF:GIF2017}. We are currently working to improve the laser-frequency noise.
 
 A strong correlation between air pressure and strain was observed in the frequency range $10^{-4} - 10^{-3}$ Hz. The small improvement achieved by correcting the strain using just the local air pressure record indicates the effect of regional air pressure onto the ground strain. This result also suggests that the use of actual strain data is crucial for baseline compensation.
 
 The main disturbance to the continuous operation of KAGRA and other ground-based GW detectors is seismic noise at low frequencies (below 1 Hz) 
 \cite{GIF:Abbott2018}. The GIF can accurately observe ground motions in that frequency range, and its signal is useful for baseline-length compensation to enhance the duty cycle of KAGRA. We have successfully demonstrated reductions in the cavity-length change in the frequency ranges of both tides and microseismic motions.

\section{Conclusion}\label{Sec:main:Conclusion}

KAGRA is a GW interferometer in Japan. 
In April 2019, the installation work was mostly completed, 
and two-week observation run called O3GK was performed in April 2020.

Calibration accuracy and detector characterization both play 
important roles in obtaining definitive results.
To evaluate the quality of the interferometer and the GW data, and to understand the interferometer environment,
physical-environment monitors and the geophysics interferometer play important roles.

For accurate calibration, two calibration instruments, PCAL and GCAL, are planned to install, with PCAL 
being used for calibration during the O3GK observations.
For reconstructing the h(t) strain, a calibration model was constructed and the calibration
parameters measured.
Three types of reconstruction pipelines were developed: online, low-latency, and 
high-latency pipelines. 
Error estimation is also important for evaluating the reconstruction pipelines and 
performing data analysis. 
Now, the high-latency h(t) strain was ready for the 
collaborators, and data analysis using them are ongoing.
For future prospect, the improvement of calibration 
accuracy and reducing the systematic error.
The combination of PCAL and GCAL will play an
important role for improvement.

As one of detector characterization activities for O3GK, the DQ vector was provided as the online process.
Because we plan to perform only offline gravitational wave searches for O3GK observational data, the cadence of providing DQ vector was not so important.
However KAGRA will perform low-latency analyses in O4 observing run.
So the framework of DQ vector production which was prepared for O3GK will be  effectively used in order to choose data segment by low-latency GW search pipelines in O4 and future observations.
The segment information which was common format with LIGO and Virgo was also provided and stored DQSEGDB in CIT.
Science segment will be used for offline GW searches by using O3GK data.
Providing more detailed segments which indicate an existence of loud glitches, is planned O4 observation.
The goal of segment production in O4 observation is used not only the data selection for GW searches but also the removal of fake events of GWs.
Noise transient investigation by Omicron and hveto is now being performed as the offline noise investigation. In the O4 observation, transient noise survey will be performed as both low-latency and offline analyses.
Results of low-latency noise investigation will be used for the veto analysis with GW search results by low-latency pipelines, improvement of contents of SummaryPage, and so on.
One of the most urgent tasks toward O4 observation for KAGRA is to reduce the cadence of noise investigation.
The data-acquisition system is integrated with the KAGRA digital control system, and 
more than 100,000 auxiliary channels were recorded with GW signals.
To evaluate the detector health and noise status, a data-quality state vector was prepared and
used to identify appropriate science segment.
The KAGRA science segments were shared with other international interferometers via DQSEGDB in a 
data server at CIT.
SummaryPages were also prepared to help identify the reason why data were flagged as``bad'' (i.e., unsuitable for GW searches).
To investigate and veto transient signals in the gravitational wave channel and auxiliary channels,
we implemented the hveto analysis technique. This technique is used in un-modeled GW searches (burst searches).
The detail investigation for 
triggered event by data analysis pipeline in the O3GK observation will be the interesting topic by using the 
rely-able tools which were described in this paper.

In one auxiliary channel, 
various types of physical environment monitors were installed 
before the O3GK observations. 
They have already helped to identify some noise sources 
and to understand their couplings to the detector sensitivity.
In future, the characterization of seismic motion at 
underground environment will be hot topic.
We are investigating the seismic motion of 
seasonal dependence and weather dependence.
Also, the seimic motion effect to interferometer by the 
earthquake, micro-seismic motion are ongoing.
Not only the seismic motion, but also the magnetic 
filed from lightening, cosmic ray and human activities 
around the experimental area are interesting topics.

%
%

The GIF is a unique feature of KAGRA. It is used to evaluate ground motions that limit the stability of the GW detector in the low-frequency region. It has been observing the actual ground motions in the KAGRA tunnel below 1 Hz with good resolution virtually continuously, with a 99.4 \% duty cycle. A strong correlation between ground motions and air pressure was found by the GIF in $10^{-4}-10^{-3}$ Hz frequency range, which cannot be estimated accurately from global models. A baseline-length compensation system for KAGRA has been successfully demonstrated using the GIF data.
This baseline-length compensation system will reduce the effect the seismic motion and will improve the duty cycle.

In this article, we focused on the introduction and history of the KAGRA calibration, detector characterization,
physical-environment monitors, and the geophysics interferometer.
Detailed results for the O3GK observations will appear in subsequent articles.


\section*{Acknowledgment}
This work was supported by MEXT, JSPS Leading-edge Research Infrastructure Program, JSPS Grant-in-Aid for Specially Promoted Research 26000005, JSPS Grant-in-Aid for Scientific Research on Innovative Areas 2905: JP17H06358, JP17H06361 and JP17H06364, JSPS Core-to-Core Program A. Advanced Research Networks, JSPS Grant-in-Aid for Scientific Research (S) 17H06133, the joint research program of the Institute for Cosmic Ray Research, University of Tokyo, National Research Foundation (NRF) and Computing Infrastructure Project of KISTI-GSDC in Korea, Academia Sinica (AS), AS Grid Center (ASGC) and the Ministry of Science and Technology (MoST) in Taiwan under grants including AS-CDA-105-M06, the LIGO project, and the Virgo project. GIF was also supported by JSPS KAKENHI Grant Number JP17H06207, the Joint Research Program of the Institute for Cosmic Ray Research (ICRR), University of Tokyo (2019-F19), and the Joint Usage/Research Center program of the Earthquake Research Institute (ERI), University of Tokyo (2019-B-03).Detchar was supported by JSPS KAKENHI Grant Number JP18K03671, PEM was supprted by JSPS KAKENHI Grant Number 19J01299 and 20H05256, the authors would like to thank Enago (www.enago.jp) for the English language review.

\end{document}